\documentclass[twocolumn,showpacs,amsmath,amssymb]{revtex4}
\begin{document}

\title[Long range interactions between alkali and alkaline-earth atoms ]{Long range interactions between alkali and alkaline-earth atoms}

\author{Jun Jiang, Yongjun Cheng and J. Mitroy}

\affiliation{School of Engineering, Charles Darwin University,
Darwin NT 0909, Australia}
\begin{abstract}
Dispersion coefficients between the alkali metal atoms (Li-Rb) and
alkaline-earth metal atoms (Be-Sr) are evaluated using matrix elements
computed from frozen core configuration interaction calculations.
Besides dispersion coefficients with both atoms in their respective
ground states, dispersion coefficients are also given for the case
where one atom is in its ground state and the other atom is in
a low lying excited state.
\end{abstract}

\maketitle

\section{Introduction}
Recently a systematic {\em ab-initio} investigation of the ground state
potential curves for the MgX dimers, where X is an alkali atom, was
reported \cite{augustovicova12a}.  The motivation for this work was the
growing interest in the production of molecules from ultracold atomic
gases \cite{doyle04a}.
Such molecules can be formed by photo-association \cite{stwalley99a,jones06a},
or by Feshbach resonance tuning \cite{kohler06a,brue12a}.  Most focus has been
on diatomic molecules consisting of two alkali atoms
\cite{fioretti98a,kraft06a,lozeille06a,zwierlein03a}.
However it has been suggested that ultracold molecules in $^2 \Sigma$ states
would be good systems for experiments on controlled chemical
reactions \cite{tscherbul06a}.  Such molecules could be formed from an alkali atom in
its ground state and an alkaline-earth atom (or Yb) in its ground state.
Recently, the vibrational spectra of CaLi and SrLi were investigated for
their sensitivity to the $m_e/M_p$ mass ratio \cite{kajita13a}.
There have been previous studies of the structure of alkali/alkaline-earth
dimers \cite{schlachta90a,bauschlicher92a,allouche94a,berry97a,bruna02a,ivanova11a}.
The most recent theoretical investigations of these molecules motivated by
cold atom physics include
LiSr \cite{guo13a}, LiYb \cite{gopakumar10a,zhang10a}, RbSr \cite{zuchowski10a} and
RbYb \cite{sorensen09a,munchow11a,lamb12a}.  A more comprehensive investigation
has been made of the LiX dimers, where X is a alkaline-earth atom
\cite{kotochigova11a,gopakumar11a}.

The present article investigates the long range interaction of various
combinations of alkali and alkaline-earth atoms.  The most efficient
method for determining the long-range interaction is by computing the
dispersion coefficients since this leads to the factorization of
one large many-body calculation into two smaller many-body calculations.
The dispersion coefficient calculations were not restricted to the
respective ground states.  Dispersion coefficients
are also given for an alkali atom in its ground state and the
$nsnp \  ^3P^o$ and  $nsnp \  ^1P^o$  excited states of the alkaline-earth
atoms.  Coefficients are given for the $nsnd \  ^{1,3}D^e$ excited
states of calcium and strontium since these $nsnd \  ^{1}D^e$ states
have a smaller excitation energy than the $nsnp \  ^1P^o$  excited states.
The dispersion coefficients for the alkaline-earth atoms in
their ground states and three of the lowest excited states of the alkali
atoms are also given.  The present work gives a comprehensive overview of the long
range interactions between the ground and the low lying excited
states of the alkali/alkaline-earth dimers.

\section{Methodology of calculation}

\subsection{Overview of van der Waals interaction calculation}

The long-range van der Waals interaction between two hetero-nuclear
atoms (i.e. the two atoms are different), with one atom in an $S$-state,
can be as a function of inter-nuclear separation, $R$
written \cite{dalgarno66a,dalgarno67a,zhang07c}
\begin{equation}
V(R) = - \frac{C_6}{R^6} - \frac{C_8}{R^8} - \frac{C_{10}}{R^{10}} - \cdots .
\end{equation}
The $C_n$ parameters are the dispersion coefficients.

The approach used to generate the dispersion coefficients is based
on the use of oscillator strength sum rules \cite{dalgarno66a,dalgarno67a}.
This reduces the calculation of the $C_n$ parameters for two
spherically symmetric atoms to summations over the products of the
absorption oscillator strengths (originating in the ground state)
divided by an energy denominator.  The sums should include
contributions from all discrete and continuum excitations.  In
practice, a pseudo-state representation is used which gives a
discrete representation of the continuum \cite{bukta74a,mitroy03f,zhang07c}.
The sum over oscillator strengths needs to be rewritten in
terms of a sum over the reduced matrix elements of the
electric multipole operator in cases where one (or both) of
the atoms is in a state with $L > 0$ \cite{zhang07c}.

The major part of any calculation involves the generation of the
lists of reduced transition matrix elements for the two atomic
states.  This involves quite lengthy calculations to generate
the excitation spectrum of the pseudo-state representation.  It
is then a relatively straightforward calculation to process the
lists of matrix elements and generate the dispersion coefficients
\cite{zhang07c,mitroy07e}.

\subsection{Structure model: The alkali atoms}

The transition arrays for the alkali atoms are essentially those
which were used in calculations of the dispersion interactions
between these atoms and the ground states of hydrogen and helium
\cite{zhang07c}.

These were computed by diagonalizing the fixed core Hamiltonian
in a large basis of Laguerre Type Orbitals (LTO).  The core
Hamiltonian is based upon a Hartree-Fock (HF) description of the
core with a semi-empirical core polarization potential
tuned to reproduce the energies of the low lying spectrum.
The oscillator strengths (and other multipole expectation values)
were computed with operators that included polarization corrections
\cite{hameed68a,hameed72a,vaeck92a,mitroy93a,mitroy03f}.

Core excitations are included in the $C_n$
calculations.  Oscillator strength distributions were
constructed by using independent estimates of the core
polarizabilities to constrain the sum rules
\cite{mitroy03e,mitroy03f,mitroy04b,zhang07a}.  The
methodology of using constrained sum rules to construct
pseudo-oscillator strength distributions has been widely
used \cite{kumar85a}.

\begin{table}[tbh]
\caption{\label{tab1}
The static dipole ($\alpha_1$) and quadrupole ($\alpha_2$) polarizabilities
(in a.u.) for the ground states of the alkali and alkaline-earth atoms.  A
recent review summarizes static dipole polarizabilities calculations and
experiments
\cite{mitroy10a}. }
\begin{ruledtabular}
\begin {tabular}{lccl}
Atom   &   CICP  & MBPT-SD  &   Other  \\
\hline
$\alpha_1$ Li     &  164.21  & 164.08 \cite{johnson08a}   &   164.11(3) (Hylleraas) \cite{tang10a} \\
                  &          &                            &     164.2(1.1) Expt. \cite{miffre06a}  \\
$\alpha_1$ Na     &  162.8   & 163.0 \cite{derevianko99a} &   162.6(3) (Hybrid) \cite{derevianko99a}\\
                  &          &                            &     162.7(8) Expt. \cite{ekstrom95a}     \\
$\alpha_1$ K      &  290.0   & 289.1 \cite{derevianko99a} &   290.2(8) (Hybrid) \cite{derevianko99a} \\
                  &          &                            &   290.8(1.4) Expt. \cite{holmgren10a}   \\
$\alpha_1$ Rb     &  315.7   & 317.4 \cite{derevianko99a} &   318.6 (Hybrid) \cite{derevianko99a}  \\
                  &          &                            &   318.8(1.4) Expt. \cite{holmgren10a}    \\
$\alpha_2$ Li     &  1,424   &  1,424(4)   \cite{porsev03a} &   1,423.26 (Hylleraas) \cite{tang09a}      \\
$\alpha_2$ Na     &  1,879   &  1,885(26)  \cite{porsev03a} &                         \\
$\alpha_2$ K      &  5,005   &  5,000(45)  \cite{porsev03a} &                         \\
                  &          &  5,018   \cite{safronova08b} &                         \\
$\alpha_2$ Rb     &  6,480   &  6,520(80)  \cite{porsev03a} &                         \\
\hline
   &  CICP  & CI+MBPT   &       Other  \\
$\alpha_1$ Be     &  37.69   & 37.76 \cite{porsev06a}     &     37.755 (ECG) \cite{komasa02a}                          \\
$\alpha_1$ Mg     &  71.35   & 71.33 \cite{porsev06a}     &     74.9(27) (Hybrid) \cite{lundin73a,miller95b}   \\
$\alpha_1$ Ca     &  159.4   & 159.0 \cite{porsev06a}     &     157.1(13) (Hybrid) \cite{porsev06a}  \\
$\alpha_1$ Sr     &  197.6   & 202.0 \cite{porsev06a}     &  197.2(2) (Hybrid) \cite{porsev06a}    \\
                  &                & 198.9 \cite{safronova13b}  &   197.14(20) (Hybrid) \cite{safronova13b}        \\
$\alpha_2$ Be     &  300.7  &  300.6(3) \cite{porsev06a} &   300.96 (ECG) \cite{komasa02a}                \\
$\alpha_2$ Mg     &  813.9  &  812(6)  \cite{porsev06a} &                              \\
$\alpha_2$ Ca     &  3,063   &  3,081(23) \cite{porsev06a}  &                             \\
$\alpha_2$ Sr     &  4,645   &  4,630(8) \cite{porsev06a}  &                       \\
\end{tabular}
\end{ruledtabular}
\end{table}

\subsection{The alkaline-earth atoms}

The use of a fixed core model reduces the calculation of the
alkaline-earths and their excited spectra to a two electron
calculation.  The two electron wavefunctions were expanded in a large
basis of two electron configurations formed from a single electron
basis mostly consisting of LTO.  Typically
the total number of one electron states would range from 150
to 200.  The use of such large basis sets means that the error
due to incompleteness of the basis is typically very small.
The semi-empirical polarization potential needs to include a
two-body term to deal with the instantaneous interaction between
the core and the two valence electrons that may be on opposite sides
of the nucleus \cite{hameed72a,mitroy88d,mitroy03f}.

Details of the calculations used to represent Be, Mg, Ca and Sr have
been previously described \cite{mitroy10d,mitroy07e,mitroy08a,mitroy08g,mitroy10b}.
We refer to these semi-empirical models of atomic structure as the
configuration interaction plus core polarization (CICP) model in subsequent
text.  The matrix element set for Sr incorporated experimental information.
An experimental value was used for the $5s^2$ $^1S^e$-$5s5p$ $^1P^o$
matrix element \cite{yasuda06a} and the energy differences for the low-lying
excitations were set to the experimental energies \cite{mitroy10b}.

\section{Results}

\subsection{Polarizabilities}

The static dipole and quadrupole polarizabilities are listed in Table \ref{tab1}.
These are defined using oscillator strength sum rules \cite{miller77a,mitroy10a},
namely
\begin{eqnarray}
\alpha_{k} & = & \sum_{n} \frac {f^{(k)}_{0n} } {\epsilon_{0n}^2} \ ,
\label{alpha1}
\end{eqnarray}
where $f^{(k)}_{0n}$ is the oscillator strength for the
$k$-order multipole operator.

\begin{table}[tbh]
\caption{\label{tab2}
The dispersion coefficients (in a.u.) for the ground state of alkali atoms
interacting with the ground states of alkaline-earth atoms. The numbers
in the square brackets denote powers of ten.  $C_6$ coefficients derived
from MBPT-SD and CI+MBPT dynamic polarizability \cite{derevianko10a} are
given in the rows with no other $C_n$ coefficients and have estimated errors
given by the numbers in brackets.  The notation $a[b]$ means $a \times 10^b$.
}
\begin{ruledtabular}
\begin{tabular}{lccc}
  &    $C_{6}$ & $C_{8}$& $C_{10}$ \\
\hline
   & \multicolumn{3}{c}{Be($2s^2$ $^1$S$^e$)} \\
Li($2s$) & 478.3    &1.743[4]  &1.038[6]      \\
         & 478(3)   &          &              \\
Na($3s$) & 522.1    &2.196[4]  &1.403[6]      \\
         & 521(4)   &          &              \\
K($4s$)  & 791.4    &4.775[4]  &3.924[6]      \\
         & 790(6)   &          &              \\
Rb($5s$) & 869.4    &5.818[4]  &5.076[6]      \\
         & 873(7)   &          &              \\
\hline
 & \multicolumn{3}{c}{Mg($3s^2$ $^1$S$^e$)}   \\
Li($2s$) & 856.8     &5.676[4]  &4.535[6]     \\
         & 853(8)    &          &           \\
Na($3s$) & 930.1     &6.727[4]  &5.735[6]    \\
         & 926(9)    &          &           \\
K($4s$)  & 1,417  & 1.282[5]  &1.333[7]  \\
         & 1,411(4)   &          &            \\
Rb($5s$) & 1,553  & 1.513[5]  &1.660[7]  \\
         & 1,556(15) &          &          \\
\hline
  &  \multicolumn{3}{c}{Ca($4s^2$ $^1$S$^e$)}        \\
Li($2s$) &  1,689          &  1.417[5]  &   1.263[7] \\
         &  1,660(14)      &            &            \\
Na($3s$) &  1,815          &  1.637[5]  &   1.564[7] \\
         &  1,782(15)      &            &             \\
K($4s$)  &  2,803          &  3.030[5]  &   3.468[7] \\
         &  2,756(23)      &            &              \\
Rb($5s$) &  3,064          &  3.535[5]  &   4.268[7] \\
         &  3,030(26)      &            &             \\
\hline
& \multicolumn{3}{c}{Sr($5s^2$ $^1$S$^e$)}           \\
Li($2s$) &   2,035        &  1.898[5]  & 1.789[7]    \\
         &   2,022(3)     &            &             \\
Na($3s$) &   2,183        &  2.172[5]  & 2.194[7]    \\
         &   2,167(4)     &            &             \\
K($4s$)  &   3,384        &  3.971[5]  & 4.765[7]    \\
         &   3,362(8)     &            &             \\
Rb($5s$) &   3,699        &  4.609[5]  & 5.833[7]    \\
         &   3,697(10)    &            &             \\
\end{tabular}
\end{ruledtabular}
\end{table}

The purpose of Table \ref{tab1} is to give an overview of the expected
accuracy of the present dispersion coefficient calculations since
polarizabilities and dispersion coefficients are computed with sum
rules using the same oscillator strengths.  Polarizabilities from high
accuracy calculations based on relativistic many body perturbation theory
\cite{safronova99a,derevianko99a,porsev03a,porsev06a,johnson08a,safronova08b,safronova13b}
are also shown in Table \ref{tab1}.  The many body theory results for one
electron atoms are an all-order relativistic many body perturbation theory
with single and double excitations (MBPT-SD) \cite{safronova99a,derevianko99a}.
For two electron atoms, many body perturbation theory is used to treat the
interaction between the core and valence electrons while the interactions between
the two valence electrons are treated with the configuration interaction
approach.  This is called the CI+MBPT approach \cite{porsev06a}.

The agreement with experiment or hybrid experimental/theoretical estimates
of the static dipole polarizability
for Li, Na and K is better than 0.5 $\%$.  The level of agreement is
$1\%$ for rubidium where relativistic effects are more important.
There is close agreement between CICP and MBPT-SD \cite{porsev03a}
quadrupole polarizabilities for the alkali atoms.

There is very good agreement between the CICP and CI+MBPT polarizabilities for
the light alkaline-earth atoms, Be and Mg.  Unfortunately, there has not been
a high precision experimental estimate of the polarizability for either of these atoms.
However, there has been one very accurate calculation of the Be polarizabilities
using an explicitly correlated gaussian (ECG) basis \cite{komasa02a}.  The ECG
polarizabilities should be correct to at least 4 digits.  Agreement with the Be
polarizabilities could hardly be better.

Additional information such as magic wavelengths and tune-out wavelengths
also give information on the polarizabilities of atomic systems
\cite{katori03a,arora07c,mitroy10a,herold12a,holmgren12a,safronova12c,jiang13a,tang13a}.
Most recently, a relativistic variant of the present calculation \cite{jiang13a} was
used to predict tune-out wavelengths for potassium.  Agreement was achieved with
the experimental value \cite{holmgren12a}.

The best experimental estimates of the polarizabilities for Ca and Sr come
from hybrid calculations where the $ns^2$ $^1S^e$-$nsnp$ $^1P^o$ matrix
elements from photo-association experiments were used to correct a
CI+MBPT calculation of the polarizability \cite{porsev06a}.  The CICP
calculation for Sr, like the hybrid calculations, used the experimental matrix
element for the resonance transition.  It is not surprising that it is in
close agreement with other hybrid theoretical/experimental estimates of the Sr
polarizability.

Table \ref{tab1} only gives ground state polarizabilities.  However, excited state
polarizabilities using the CICP method have been given
\cite{zhang07a,zhang07c,mitroy10d,mitroy07e,mitroy08a,mitroy08g,mitroy10b}.
The polarizabilities of potassium serve as an indicative example.  CICP calculations
gave 615 a.u. for K($4p$), 4997 a.u. for K($5s$), and 1419 a.u. for K($3d$) \cite{zhang07c}.
MBPT-SD calculations give 611 a.u. for K($4p_{1/2}$), 620 a.u. for K($4p_{3/2}$),
4961 a.u. for K($5s_{1/2}$), 1420 a.u. for K($3d_{3/2}$), and 1412 a.u. for
K($3d_{5/2}$) \cite{safronova13a}.
The evaluation of the oscillator strength sum-rules in previous CICP investigations
often used calculated energy differences.  While this has only a small effect on the
ground state polarizabilities, the use of calculated energy differences will
introduce larger differences when applied to excited states where the
energy differences are smaller and thereby making the polarizabilities
more sensitive to small errors in the calculated energies.  Spin-orbit
energy splittings can also impact polarizabilities of the excited
states for the heavier atoms.  For example, the static dipole polarizabilities
of the $5p_{J}$ spin-orbit doublet differ by 8$\%$ \cite{safronova11b}
and the present dispersion coefficients should be interpreted as the
average for a spin-orbit doublet or triplet state.

\begin{table*}
\caption{\label{tab3}The dispersion coefficients (in a.u.) for the ground state of alkali atoms
 interacting with the $nsnp$ $^1P^o$ and $^3P^o$  excited states of alkaline-earth atoms.
$n=2, 3, 4$ and $5$ for Be, Mg, Ca and Sr, respectively.
The notation $a[b]$ means $a \times 10^b$.   Dispersion coefficients which
could be influenced by an accidental degeneracy with pseudo-states in the
alkali atom continuum are indicated by underlining.  }
\begin{ruledtabular}
\begin{tabular}{lcccccccc}
       && \multicolumn{3}{c}{$nsnp$ $^1P^o$} & & \multicolumn{3}{c}{$nsnp$ $^3P^o$}   \ \\
\cline{3-5}
\cline{7-9}
        &&   $C_6$    &   $C_8$   &   $C_{10}$                      && $C_6$  & $C_8$ &$C_{10}$   \ \\
\hline
  & &\multicolumn{7}{c}{Be} \\
Li$(2s)$  & $\Sigma$ &      1.228[3]     &      2.370[4]     &     4.737[7]  &&      4.812[2]     &      1.751[4]     &     1.043[6]   \\
          & $\Pi$    &      9.516[2]     &      2.725[4]     &     1.882[7]  &&      4.930[2]     &      1.785[4]     &     1.061[6]   \\
Na$(3s)$  & $\Sigma$ &      1.176[3]     &   $-$3.650[3]     &     1.303[7]  &&      5.248[2]     &      2.206[4]     &     1.410[6]   \\
          & $\Pi$    &      9.883[2]     &      2.335[4]     &     6.321[6]  &&      5.377[2]     &      2.250[4]     &     1.435[6]   \\
K$(4s)$   & $\Sigma$ &  \underline{2.173[3]}  & \underline{7.630[4]}  & \underline{$-$5.181[6]}  &&      7.970[2]     &      4.807[4]     &     3.948[6]   \\
          & $\Pi$    &  \underline{1.611[3]}  & \underline{8.056[4]}  & \underline{2.163[6]}  &&      8.161[2]     &      4.902[4]     &     4.017[6]  \\
Rb$(5s)$  & $\Sigma$ &  \underline{2.444[3]}  & \underline{9.739[4]} &  \underline{5.600[6]}  &&      8.756[2]     &      5.861[4]     &     5.109[6] \\
          & $\Pi$    &  \underline{1.778[3]}  & \underline{1.002[5]} &  \underline{7.443[6]}  &&      8.964[2]     &      5.976[4]     &     5.199[6] \\
\\ \hline
   &&\multicolumn{7}{c}{Mg } \\
Li$(2s)$  & $\Sigma$ &      2.310[3]     &     7.052[5] &     1.340[8]  &&      1.244[3]     &     2.598[5] &     2.431[7] \\
          & $\Pi$    &      1.835[3]     &     7.147[4] &     1.472[7]  &&      1.063[3]     &     5.923[4] &     3.840[6] \\
Na$(3s)$  & $\Sigma$ &      2.239[2]     &     4.099[5] &     8.929[7]  &&      1.337[3]     &     2.814[5] &     2.925[7] \\
          & $\Pi$    &      1.396[3]     &  $-$2.585[4] &  $-$5.971[6]  &&      1.148[3]     &     6.921[4] &     5.053[6] \\
K$(4s)$   & $\Sigma$ & \underline{4.148[3]} & \underline{1.300[6]} & \underline{1.967[8]}  &&      2.060[3]     &     4.879[5] &     6.091[7] \\
          & $\Pi$    & \underline{3.131[3]} & \underline{2.038[5]} & \underline{4.813[6]}  &&      1.760[3]     &     1.443[5] &     1.330[7] \\
Rb$(5s)$  & $\Sigma$ & \underline{4.629[3]} & \underline{1.537[6]} & \underline{2.606[8]} &&      2.252[3]     &     5.489[5] &     7.382[7] \\
          & $\Pi$    & \underline{3.442[3]} & \underline{2.783[5]} & \underline{1.870[7]}  &&      1.926[3]     &     1.719[5] &     1.708[7] \\
\\ \hline
   &&\multicolumn{7}{c}{Ca} \\
Li$(2s)$  & $\Sigma$ &   $-$5.557[2]     &     1.082[6] &     1.515[8]  &&      2.307[3]     &     6.241[5] &     6.669[7]\\
          & $\Pi$    &      1.474[3]     &     2.310[5] &     1.714[7]  &&      2.065[3]     &     1.205[5] &     8.890[6]\\
Na$(3s)$  & $\Sigma$ &   $-$2.476[3]     &     1.281[6] &     1.742[8]  &&      2.438[3]     &     6.652[5] &     7.864[7]\\
          & $\Pi$    &      1.096[3]     &     3.046[5] &     2.209[7]  &&      2.188[3]     &     1.405[5] &     1.145[7] \\
K$(4s)$   & $\Sigma$ &      1.164[3]     &  $-$9.103[4] &     3.670[8]  &&      3.871[3]     &     1.151[6] &     1.590[8]\\
          & $\Pi$    &      2.983[3]     &  $-$1.946[5] &     6.467[7]  &&      3.461[3]     &     2.886[5] &     2.884[7] \\
Rb$(5s)$  & $\Sigma$ &      4.641[3]     &     8.794[5] &     4.611[8]  &&      4.223[3]     &     1.287[6] &     1.908[8]\\
          & $\Pi$    &      4.099[3]     &     1.348[5] &     9.109[7]  &&      3.778[3]     &     3.442[5] &     3.664[7]\\
\hline
   &&\multicolumn{7}{c}{Sr} \\
Li$(2s)$  & $\Sigma$ &   $-$1.753[3]     &     1.408[6] &     2.240[8]  &&      3.116[3]     &     9.286[5] &     1.086[8]\\
          & $\Pi$    &      1.560[3]     &     2.695[5] &     2.235[7]  &&      2.718[3]     &     1.575[5] &     1.257[7]\\
Na$(3s)$  & $\Sigma$ &   $-$5.563[3]     &     1.589[6] &     2.541[8]  &&      3.270[3]     &     9.838[5] &     1.266[8]\\
          & $\Pi$    &      7.231[2]     &     3.403[5] &     2.831[7]  &&      2.860[3]     &     1.842[5] &     1.602[7] \\
K$(4s)$   & $\Sigma$ &      3.002[2]     &  $-$2.382[7] &     5.054[8]  &&      5.257[3]     &     1.703[6] &     2.517[8]\\
          & $\Pi$    &      3.428[3]     &  $-$8.213[7] &     7.551[7]  &&      4.582[3]     &     3.769[5] &     3.962[7] \\
Rb$(5s)$  & $\Sigma$ &      1.068[3]     &     4.886[5] &     6.078[8]  &&      5.735[3]     &     1.899[6] &     3.002[8] \\
          & $\Pi$    &      3.927[3]     &  $-$1.103[5] &     9.973[7]  &&      5.000[3]     &     4.503[5] &     5.009[7] \\
\end{tabular}
\end{ruledtabular}
\end{table*}
\subsection{Ground state dispersion coefficients}

Table \ref{tab2} lists the $C_6$, $C_8$ and $C_{10}$ dispersion coefficients
between all combinations consisting of a ground state alkali atom and
a ground state alkaline-earth atom. There have been two previous comprehensive tabulations
of $C_n$ coefficients for these combinations of atoms.  The tabulation
by Standard and Certain \cite{standard85a} can be regarded as
obsolete \cite{mitroy03f}. More recently, dynamic polarizabilities from MBPT-SD
and CI+MBPT calculations have been used to estimate dispersion
coefficients for many alkali/alkaline-earth dimers \cite{derevianko10a}.
However these calculations were restricted to the lowest order $C_6$
coefficients.

For all practical purposes, the present $C_6$ parameters and the MBPT based
$C_6$ coefficients are identical for dimers containing Li, Na, K, Rb, Be
and Mg.  There is not a single instance where the two sets of calculations
differ by more than $1\%$.  There is also better than $1\%$ agreement between
the CICP and CI+MBPT dispersion coefficients for strontium.  The largest
differences occur for the dimers involving calcium, where the CICP $C_6$
is just over 1$\%$ larger than the CI+MBPT values which use an experimental
matrix element for the resonance transition.  This $C_6$ difference was expected since
the CICP values of $\alpha_1$ for calcium were just over 1$\%$ larger than the
hybrid CI+MBPT calculation of $\alpha_1$.

\subsection{Dispersion coefficients for the alkaline-earth excited states}

Table \ref{tab3} details the dispersion coefficients for the $nsnp$
$^{1,3}P^o$ alkaline-earth excited states.  In the case of calcium and
strontium,
the first singlet excited state is actually the $ns(n\!-\!1)d$ $^1D^e$ state.
For purposes of completeness, dispersion coefficients between the
$ns(n\!-\!1)d$ $^{1,3}D^e$ states states of calcium and strontium with the
alkali ground states are listed in Table \ref{tab4}.

The presence of a downward
transition for the $^1P^o$ excited state makes it possible for the
dispersion coefficients to be negative, thereby indicating a repulsive
dispersion interaction.  Contributions of the $C_n$ coefficients can be
negative when the total energy of the transitions originating from the
two atoms is negative.  The $^3P^o$ state does not have a spin-allowed
transition to the ground state so all the dispersion coefficients are positive.
Examples of a negative $C_6$ coefficient occur for some of the $\Sigma$
states in Table \ref{tab3}.

\begin{table*}[tbh]
\caption{\label{tab4}The dispersion coefficients (in a.u.) for the ground state of alkali atoms
interacting with the $ns(n\!-\!1)d$ $^1D^e$ and $^3D^e$  excited states of
calcium and strontium.
The notation $a[b]$ means $a \times 10^b$. }
\begin{ruledtabular}
\begin{tabular}{lcccccccc}
       && \multicolumn{3}{c}{$ns(n\!-\!1)d$ $^1D^e$} & & \multicolumn{3}{c}{$ns(n\!-\!1)d$ $^3D^e$}   \ \\
\cline{3-5}
\cline{7-9}
        &&   $C_6$    &   $C_8$   &   $C_{10}$                      && $C_6$  & $C_8$ &$C_{10}$   \ \\
\hline
     & & &   \multicolumn{5}{c}{Ca}   \ \\
Li$(2s)$  & $\Sigma$ &      1.605[3]     &     1.347[5] &     4.943[7] &&      3.033[3]     &     2.852[5] &     3.749[7] \\
          & $\Pi$    &      1.673[3]     &     1.356[5] &     1.897[7] &&      2.806[3]     &     1.386[5] &     1.404[7] \\
          & $\Delta$ &      1.876[3]     &     1.357[5] &     1.258[7] &&      2.124[3]     &     1.203[5] &     1.080[7] \\
Na$(3s)$  & $\Sigma$ &      1.713[3]     &     7.834[4] &     5.944[7] &&      2.988[3]     &     3.052[5] &     4.215[7] \\
          & $\Pi$    &      1.783[3]     &     1.316[5] &     2.451[7] &&      2.773[3]     &     1.638[5] &     1.684[7] \\
          & $\Delta$ &      1.993[3]     &     1.567[5] &     1.562[7] &&      2.127[3]     &     1.392[5] &     1.315[7] \\
K$(4s)$   & $\Sigma$ &      2.679[3]     &     3.577[5] &  $-$4.121[8] &&      5.330[3]     &     6.100[5] &     8.687[7] \\
          & $\Pi$    &      2.792[3]     &     3.153[5] &  $-$1.809[8] &&      4.906[3]     &     3.328[5] &     3.978[7] \\
          & $\Delta$ &      3.133[3]     &     3.014[5] &     2.023[7] &&      3.632[3]     &     2.861[5] &     3.095[7] \\
Rb$(5s)$  & $\Sigma$ &      2.927[3]     &     4.285[5] &     3.549[7] &&      5.840[3]     &     7.103[5] &     1.036[8] \\
          & $\Pi$    &      3.050[3]     &     3.726[5] &     2.090[7] &&      5.373[3]     &     4.029[5] &     4.920[7] \\
          & $\Delta$ &      3.419[3]     &     3.544[5] &     4.026[7] &&      3.972[3]     &     3.434[5] &     3.845[7] \\
\hline
     &  &  &  \multicolumn{5}{c}{Sr}   \ \\
Li$(2s)$  & $\Sigma$ &      1.996[3]     &     1.228[5] &     8.498[7]   &&      3.605[3]     &     4.384[5] &     6.361[7]\\
          & $\Pi$    &      2.057[3]     &     1.490[5] &     2.591[7]   &&      3.312[3]     &     1.918[5] &     1.908[7] \\
          & $\Delta$ &      2.239[3]     &     1.658[5] &     1.605[7]   &&      2.433[3]     &     1.435[5] &     1.275[7] \\
Na$(3s)$  & $\Sigma$ &      2.125[3]     &  $-$9.792[4] &     9.931[7]   &&      3.686[3]     &     4.737[5] &     7.180[7]\\
          & $\Pi$    &      2.189[3]     &     9.393[4] &     3.324[7]   &&      3.401[3]     &     2.288[5] &     2.336[7] \\
          & $\Delta$ &      2.380[3]     &     1.908[5] &     1.983[7]   &&      2.545[3]     &     1.698[5] &     1.594[7] \\
K$(4s)$   & $\Sigma$ &      3.342[3]     &     4.544[5] &     3.521[8]   &&      6.271[3]     &     8.915[5] &     1.418[8] \\
          & $\Pi$    &      3.443[3]     &     3.853[5] &     1.453[8]   &&      5.739[3]     &     4.415[5] &     5.377[7] \\
          & $\Delta$ &      3.746[3]     &     3.669[5] &     4.833[7]   &&      4.144[3]     &     3.395[5] &     3.666[7] \\
Rb$(5s)$  & $\Sigma$ &      3.653[3]     &     5.536[5] &  $-$1.951[8]   &&      6.861[3]     &     1.025[6] &     1.674[8] \\
          & $\Pi$    &      3.762[3]     &     4.591[5] &  $-$9.572[7]   &&      6.279[3]     &     5.286[5] &     6.642[7] \\
          & $\Delta$ &      4.090[3]     &     4.309[5] &     4.238[7]   &&      4.531[3]     &     4.063[5] &     4.562[7] \\
\end{tabular}
\end{ruledtabular}
\end{table*}

Table \ref{tab3} exhibits some expected trends.  The $C_6$ coefficients
tend to increase as the alkali atoms get larger in the Li $\to$ Rb sequence.
This is expected since the polarizabilities increase from Li $\to$ Rb. There
is also an increase in $C_6$ for the $nsnp$ $^3P^o$ states as the atoms
increase in size from Be $\to$ Sr.   This is again a polarizability related
increase.   A steady increase in $nsnp$ $^1P^o$ state $C_6$ values does not
occur as the atomic size increases from Mg $\to$ Sr.  These states  have
downward transitions and the Mg $3s3p$ $^1S^e$ polarizability \cite{mitroy07e}
is about twice the size of the Sr $5s5p$ $^1P^o$ polarizability \cite{mitroy10b}.

There are a number of apparent irregularities when examining the $C_n$ values for
a sequence of atoms, e.g from Li $\to$ K, or Be $\to$ Sr in Table \ref{tab3}
and the later tables.  This occurs because the energy denominators in the
sum rules for the dispersion coefficients now have single atom excitation
energies that can be both positive and negative.  For example,
negative $C_6$ values occur for some $nsnp$ $^1P^o$ states.  What has occurred
is that the energy of the downward transition of the alkaline-earth exceeds the
energy increase of some of the upward transitions of the alkali atoms.  In addition,
the energies of the downward transition and the upward transition were nearly equal,
so the energy denominator in the sum-rule
was small, thereby enhancing the contributions from these terms.   There were
also some negative $C_8$ and $C_{10}$ values for some dimers.  For the most part,
these were also found to be caused by a near-zero in the $C_n$ sum-rules caused
by the near cancellation of an energy increasing transition (and not necessarily
a dipole transition) of the alkali atom and the energy decreasing transition of
the $nsnp$ $^1P^o$ state.  In some cases, accidental near-degeneracies in the
energy denominator leads to dispersion coefficients for one dimer that seem to
bear little relation to those of another dimer for which one would expect similar
dispersion coefficients.  As a specific example, one can refer to the negative $C_8$
coefficients for the Ca($4s4p$\ $^1P^o$)-K($4s$) dimer. These were caused by the
Ca($4s4p$\ $^1P^o$) de-excitation energy of 0.107 a.u. being very close to
and larger than the K($4s \to 5p$) excitation energy of 0.096 a.u.  Something
similar occurs for the $C_8$ coefficients of the Sr($5s5p$\ $^1P^o$)-K($5s$) dimer.

There is one potential problem with some of the dispersion coefficients
involving excited states when the de-excitation energies from the
alkaline-earth excited states are larger
than the ionization energies of the alkali atoms.  Formally, the energy denominator
in the perturbation theory sum-rules for these combinations would have a
zero arising when the excitation energy in the alkali atom continuum is equal to
the de-excitation energy of the $nsnp$ $^1P^o$ excited states.  The dispersion
interaction in this case will have an imaginary part.  There is also the possibility
that an accidental near zero energy between the $nsnp$ $^1P^o$ de-excitation energy and
the energy of the one of the pseudo-states in the alkali atom pseudo-continua could
lead to an error in the calculation of the dispersion coefficients.  The only tabulated
$C_n$ coefficients susceptible to this problem occur in Table \ref{tab3}.
The dimers involved contain either the K and Rb atoms interacting with the
Be($2s2p$ \ $^1P^o$) and Mg($3s3p$ \ $^1P^o$) states and the dispersion coefficients
that may be susceptible to this problem are underlined in Table \ref{tab3}.

\begin{table*}[th]
\caption{\label{tab5}The dispersion coefficients (in a.u.) for the excited states of alkali atoms
interacting with the ground states of alkaline-earth.
The notation $a[b]$ means $a \times 10^b$. }
\begin{ruledtabular}
\begin{tabular}{lcccccccc}
                     &         &  $C_{6}$ & $C_{8}$  & $C_{10}$ &         & $C_{6}$  & $C_{8}$ & $C_{10}$ \\
\hline
                     & \multicolumn{4}{c}{Li($2p$)}             & \multicolumn{4}{c}{Li($3s$)}\\
Be($2s^2$ $^1S^e$) &$\Sigma$ &1.351[3] &1.653[5] &1.675[7] && 4.228[3]  &1.015[6] &2.975[8]  \\
                     &$\Pi$    &656.4    &1.825[3] &8.106[4] &&           &          &     \\
Mg($3s^2$ $^1S^e$) &$\Sigma$ &2.630[3] &3.647[5] &4.751[7] && 7.961[3] &2.101[6] &6.528[8]  \\
                     &$\Pi$    &1.224[3] &4.137[4] &2.227[6] &&           &          &     \\
Ca($4s^2$ $^1S^e$) &$\Sigma$ &7.117[3] &9.060[5] &1.184[8] && 1.786[4] &4.809[6] &1.504[9]  \\
                     &$\Pi$    &2.880[3] &1.611[5] &8.725[6] &&           &    &     \\
Sr($5s^2$ $^1S^e$) &$\Sigma$ &9.710[3] &1.254[6] &1.594[8] && 2.231[4] &6.134[6] &1.934[9]  \\
                     &$\Pi$    &3.751[3] &2.589[5] &1.385[7] &&           &    &     \\
\hline
                     & \multicolumn{4}{c}{Na($3p$)}             & \multicolumn{4}{c}{Na($4s$)}\\
Be($2s^2$ $^1S^e$) &$\Sigma$ &2.064[3] &3.529[5] &4.665[7] && 4.571[3] &1.178[6] &3.679[8]  \\
                     &$\Pi$    &1.043[3] &4.350[3] &2.694[5] &&     &    &     \\
Mg($3s^2$ $^1S^e$) &$\Sigma$ &4.032[3] &7.437[5] &1.200[8] && 8.605[3] &2.424[6] &7.992[8]  \\
                     &$\Pi$    &1.957[3] &6.733[4] &4.213[6] &&     &    &     \\
Ca($4s^2$ $^1S^e$) &$\Sigma$ &1.148[4] &1.814[6] &2.903[8] && 1.922[4] &5.531[6] &1.836[9]  \\
                     &$\Pi$    &4.764[3] &2.686[5] &1.558[7] &&     &    &     \\
Sr($5s^2$ $^1S^e$) &$\Sigma$ &1.647[4] &2.539[6] &3.848[8] && 2.390[4] &7.032[6] &2.357[9]  \\
                     &$\Pi$    &6.405[3] &4.586[5] &2.437[7] &&     &    &     \\
\hline
                     & \multicolumn{4}{c}{K($4p$)}              & \multicolumn{4}{c}{K($5s$)}\\
Be($2s^2$ $^1S^e$) &$\Sigma$ &2.654[3] &5.953[5] &9.536[7] && 5.861[3] &1.901[6] &7.307[8]  \\
                     &$\Pi$    &1.413[3] &9.000[3] &6.306[5] &&     &    &     \\
Mg($3s^2$ $^1S^e$) &$\Sigma$ &5.066[3] &1.222[6] &2.320[8] && 1.103[4] &3.846[6] &1.544[9]  \\
                     &$\Pi$    &2.626[3] &9.578[4] &6.898[6] &&     &    &     \\
Ca($4s^2$ $^1S^e$) &$\Sigma$ &1.231[4] &2.791[6] &5.515[8] && 2.453[4] &8.721[6] &3.521[9]  \\
                     &$\Pi$    &5.840[3] &3.297[5] &2.393[7] &&     &    &     \\
Sr($5s^2$ $^1S^e$) &$\Sigma$ &1.593[4] &3.602[6] &7.234[8] && 3.045[4] &1.103[7] &4.489[9]  \\
                     &$\Pi$    &7.336[3] &4.978[5] &3.665[7] &&     &    &     \\
\hline
                     & \multicolumn{4}{c}{Rb($5p$)}              & \multicolumn{4}{c}{Rb($6s$)}\\
Be($2s^2$ $^1S^e$) &$\Sigma$ &2.988[3] &7.496[5] &1.311[8] && 6.280[3] &   2.163[6] &    8.788[8]  \\
                     &$\Pi$    &1.620[3] &1.272[4] &9.225[5] &&          &            &             \\
Mg($3s^2$ $^1S^e$) &$\Sigma$ &5.687[3] &1.524[6] &3.119[8] && 1.181[4] &   4.359[6] &    1.845[9]  \\
                     &$\Pi$    &3.009[3] &1.135[5] &8.725[6] &&          &            &            \\
Ca($4s^2$ $^1S^e$) &$\Sigma$ &1.363[4] &3.457[6] &7.367[8] && 2.624[4] &   9.871[6] &    4.200[9]  \\
                     &$\Pi$    &6.654[3] &3.797[5] &2.940[7] &&          &            &             \\
Sr($5s^2$ $^1S^e$) &$\Sigma$ &1.753[4] &4.431[6] &9.625[8] && 3.256[4] &   1.247[7] &    5.348[9]  \\
                     &$\Pi$    &8.331[3] &5.681[5] &4.460[7] &&          &            &             \\
\end{tabular}
\end{ruledtabular}
\end{table*}

While the $nsnp$ $^1P^o$ state is the lowest excited state for Be and Mg, it is
not the lowest energy excited state for Ca and Sr.  In these cases, the
$4s3d$ $^1D^e$ and $5s4d$ $^1D^e$ states are the lowest energy excited states for
Ca and Sr respectively.  Dispersion coefficients for the $^1D^e$ and $^3D^e$
states are listed in Table \ref{tab4}.  The $C_6$ coefficients for the
$^1D^e$ states are all positive since these states do not have an energy
decreasing dipole transition.  Some of the $C_{10}$ coefficients are
also negative, this occurs because the transition energies of the
$ns \to (n\!-\!1)d$ states are almost the same as transition energies
of the $ns(n\!-\!1)d \to ns^2$ transition energies.  This leads to the
energy denominator in the sum-rule used to compute $C_{10}$
\cite{zhang07c} being close to zero.
The negative $C_8$ coefficient for the Na-Sr dimer arises from the near-equality
of the $3s \to 3p$ and $5s4d \ ^1D^e \to 5s^2 \ ^1S^e$ transition energies.

All the dispersion coefficients involving the alkaline-earth $ns(n\!-\!1)d \ ^3D^e$
states are positive.  This state has an energy decreasing dipole transition to
the $nsnp \ ^3P^o$ state.  However, the transition energy for this transition
is very small and there are no accidental near-equalities in energies with any
transitions emanating from the alkali ground states.

\begin{table*}[th]
\caption{\label{tab6}The dispersion coefficients (in a.u.) for the lowest $nd$ states of alkali atoms
interacting with the ground states of the alkaline-earth atoms. The numbers
in the square brackets denote powers of ten.}
\begin{ruledtabular}
\begin{tabular}{lcccccccc}
                     &         &  $C_{6}$ & $C_{8}$  & $C_{10}$ &         & $C_{6}$  & $C_{8}$ & $C_{10}$ \\
\hline
                     & \multicolumn{4}{c}{Li($3d$)}             & \multicolumn{4}{c}{Na($3d$)}\\
Be($2s^2$ $^1S^e$) &$\Sigma$ &      5.972[3]     &     4.208[6] &     2.273[9] &&      5.849[3]     &     4.073[6] &     2.176[9] \\
                     &$\Pi$    &      5.274[3]     &     1.408[6] &     8.463[7] &&      5.167[3]     &     1.362[6] &     8.076[7] \\
                     &$\Delta$ &      3.181[3]     &  $-$2.547[5] &  $-$2.674[6] &&      3.120[3]     &  $-$2.462[5] &  $-$2.577[6] \\
Mg($3s^2$ $^1S^e$) &$\Sigma$ &      1.137[4]     &     8.323[6] &     4.581[9] &&      1.110[4]     &     8.015[6] &     4.389[9] \\
                     &$\Pi$    &      1.002[4]     &     2.905[6] &     3.147[8] &&      9.788[3]     &     2.801[6] &     3.026[8] \\
                     &$\Delta$ &      5.984[3]     &  $-$2.947[5] &  $-$1.204[7] &&      5.857[3]     &  $-$2.815[5] &  $-$1.152[7] \\
Ca($4s^2$ $^1S^e$) &$\Sigma$ &      2.711[4]     &     1.837[7] &     1.033[10]&&      2.564[4]     &     1.747[7] &     9.880[9] \\
                     &$\Pi$    &      2.374[4]     &     6.338[6] &     8.662[8] &&      2.250[4]     &     6.076[6] &     8.298[8] \\
                     &$\Delta$ &      1.365[4]     &  $-$3.631[5] &  $-$3.683[7] &&      1.307[4]     &  $-$3.435[5] &  $-$3.379[7] \\
Sr($5s^2$ $^1S^e$) &$\Sigma$ &      3.674[4]     &     2.416[7] &     1.343[10]&&      3.361[4]     &     2.283[7] &     1.282[10] \\
                     &$\Pi$    &      3.206[4]     &     8.288[6] &     1.221[9] &&      2.943[4]     &     7.943[6] &     1.168[9] \\
                     &$\Delta$ &      1.805[4]     &  $-$2.914[5] &  $-$5.324[7] &&      1.690[4]     &  $-$2.835[5] &  $-$4.573[7] \\
\hline
                     & \multicolumn{4}{c}{K($3d$)}              & \multicolumn{4}{c}{Rb($4d$)}\\
Be($2s^2$ $^1S^e$) &$\Sigma$ &      4.147[3]     &     2.340[6] &     1.039[9]  &&  3.327[3] &   1.646[6] &    6.464[8]  \\
                     &$\Pi$    &      3.664[3]     &     7.704[5] &     3.625[7]  &&  2.943[3] &   5.373[5] &    2.166[7] \\
                     &$\Delta$ &      2.215[3]     &  $-$1.396[5] &  $-$1.272[6]  &&  1.791[3] &$-$9.586[4] & $-$8.224[5] \\
Mg($3s^2$ $^1S^e$) &$\Sigma$ &      7.833[3]     &     4.623[6] &     2.125[9]  &&  6.250[3] &   3.271[6] &    1.337[9]  \\
                     &$\Pi$    &      6.907[3]     &     1.613[6] &     1.562[8]  &&  5.516[3] &   1.143[6] &    1.034[8]\\
                     &$\Delta$ &      4.129[3]     &  $-$1.318[5] &  $-$5.260[6]  &&  3.315[3] &$-$7.370[4] & $-$2.896[6]  \\
Ca($4s^2$ $^1S^e$) &$\Sigma$ &      1.778[4]     &     1.179[7] &     5.300[9]  &&  1.397[4] &   7.951[6] &    3.094[9]  \\
                     &$\Pi$    &      1.560[4]     &     4.092[6] &     6.629[8]  &&  1.227[4] &   2.782[6] &    3.263[8] \\
                     &$\Delta$ &      9.046[3]     &  $-$9.458[4] &  $-$1.197[5]  &&  7.150[3] &$-$6.966[3] & $-$4.764[6]  \\
Sr($5s^2$ $^1S^e$) &$\Sigma$ &      2.226[4]     &     3.871[7] &     5.877[9]  &&  1.740[4] &   1.081[7] &    4.209[9]  \\
                     &$\Pi$    &      1.950[4]     &     1.317[7] &     5.303[8]  &&  1.526[4] &   3.822[6] &    5.827[8] \\
                     &$\Delta$ &      1.121[4]     &     1.211[4] &  $-$2.326[7]  &&  8.825[3] &   9.168[4] &    3.038[6]  \\
\end{tabular}
\end{ruledtabular}
\end{table*}

\subsection{Dispersion coefficients for the alkali excited states}

Table \ref{tab5} gives the dispersion coefficients for the $np$ and
$(n\!+\!1)s$ alkali states interacting with the ground states of the
alkaline earth atoms.  Table \ref{tab6} gives the dispersion coefficients
between the alkaline-earth ground states and the lowest alkali $nd$ states.
All the coefficients are positive.

All the $C_n$ coefficients in Table \ref{tab5} are positive and obey
predictable trends. The coefficients get bigger as the alkaline-earth
atoms change from beryllium to strontium.  The coefficients also increase
as the alkali atoms increase in size from lithium to rubidium.

The $C_n$ coefficients in Table \ref{tab6} also obey regular trends.
One trend is for the $C_6$ coefficients to decrease as the alkali
atoms increase in size from lithium to rubidium.  This might seem
counterintuitive but the polarizabilities of the lowest alkali $nd$
states tend to decrease in size from sodium to rubidium \cite{zhang07c}.
There is also a trend for the $C_6$ values to increase in size for the
larger alkaline-earth atoms.

The $C_8$ and $C_{10}$  coefficients are negative for the dimers
with $\Delta$ symmetry.  This is not the consequence of transitions
that decrease energy.   Rather, the equations \cite{marinescu99a,zhang07c}
for $C_n$ coefficients of $\Delta$ symmetry allow the possibility
that the dispersion coefficients can be negative.

\section{Conclusions}

Large scale CI calculations have been used to generate polarizabilities
and dispersion coefficients for many combinations of the alkali and
alkaline-earth atoms.   The underlying accuracy of the calculation is
known by reference to previous calculations using the same structure
model \cite{mitroy03f,mitroy10d,mitroy07e,mitroy08a,mitroy08g,mitroy10b}.

The most important coefficients are those for the alkali and alkaline-earth
atoms in their respective ground states. These coefficients should be
accurate to close to 1$\%$ for many combinations and a reasonable upper
limit of the maximum error would be 2-3$\%$. Somewhat surprisingly, the
only previous tabulation of the full $C_6$, $C_8$ and $C_{10}$ set
for these dimers was the Standard and Certain
compilation \cite{standard85a} and the present $C_n$ values in many
cases are an order of magnitude more precise (an assessment of the
accuracy of Standard and Certain compilation for dimers involving
alkaline-earth atoms has already been published \cite{mitroy03f}).
The same degree of precision is not present for dispersion coefficients
in dimers containing excited states.  The excited state calculations involve
de-exciting transitions that can lead to cancellations in the sum-rules
used for the computation of the $C_n$ coefficients.  In these cases,
the uncertainties in the dispersion coefficients can easily exceed
10$\%$.  These cases can be identified by looking for anomalies in
pattern of $C_n$ coefficients during an examination over a group of similar
atoms.

One limitation of the present calculation is the absence of the
spin-orbit interaction and the use of $LS$ coupling.
As mentioned earlier, the difference in the
polarizabilities of the Rb($5p$) spin-orbit doublet is 8$\%$.  This
would then translate into similar differences in the $C_6$ coefficients
involving this state.  Similarly, the spin-orbit energy splitting for
the $5s5p$ $^3P^o$ level of Sr is about 0.001 Hartree.  Given the close
proximity to the $5s4d$ $^3D^e$ state which has a binding energy that
is only 0.016 a.u. smaller, one could easily see the polarizabilities of
the $^3P^o$ spin-orbit states differing by 10$\%$, again leading to a
difference of 10$\%$ in the $C_6$ values.  The possible impact of spin-orbit
effects was a primary factor in deciding not to extend the calculations
of the heavier cesium and barium atoms.  The present results are best
regarded as giving a set of average dispersion coefficients.
The differences in dispersion coefficients involving spin-orbit doublets
could be magnified if the sum rules contain near degeneracies in some
of the energy denominators.  Taking the
present calculations of the dispersion coefficients to the next level of
accuracy would require properly relativistic structure calculations.

\section{Acknowledgments}

This work was supported in part by the Australian Research
Council Discovery Project, DP-1092620.



\begin{thebibliography}{76}
\expandafter\ifx\csname natexlab\endcsname\relax\def\natexlab#1{#1}\fi
\expandafter\ifx\csname bibnamefont\endcsname\relax
  \def\bibnamefont#1{#1}\fi
\expandafter\ifx\csname bibfnamefont\endcsname\relax
  \def\bibfnamefont#1{#1}\fi
\expandafter\ifx\csname citenamefont\endcsname\relax
  \def\citenamefont#1{#1}\fi
\expandafter\ifx\csname url\endcsname\relax
  \def\url#1{\texttt{#1}}\fi
\expandafter\ifx\csname urlprefix\endcsname\relax\def\urlprefix{URL }\fi
\providecommand{\bibinfo}[2]{#2}
\providecommand{\eprint}[2][]{\url{#2}}

\bibitem[{\citenamefont{{Augustovi{\v c}ov{\'a}} and
  {Sold{\'a}n}}(2012)}]{augustovicova12a}
\bibinfo{author}{\bibfnamefont{L.}~\bibnamefont{{Augustovi{\v c}ov{\'a}}}}
  \bibnamefont{and}
  \bibinfo{author}{\bibfnamefont{P.}~\bibnamefont{{Sold{\'a}n}}},
  \bibinfo{journal}{J.~Chem.~Phys.} \textbf{\bibinfo{volume}{136}},
  \bibinfo{pages}{084311} (\bibinfo{year}{2012}).

\bibitem[{\citenamefont{Doyle et~al.}(2004)\citenamefont{Doyle, Friedrich,
  Krems, and Masnou-Seeuws}}]{doyle04a}
\bibinfo{author}{\bibfnamefont{J.}~\bibnamefont{Doyle}},
  \bibinfo{author}{\bibfnamefont{B.}~\bibnamefont{Friedrich}},
  \bibinfo{author}{\bibfnamefont{R.~V.} \bibnamefont{Krems}}, \bibnamefont{and}
  \bibinfo{author}{\bibfnamefont{F.}~\bibnamefont{Masnou-Seeuws}},
  \bibinfo{journal}{Eur.~Phys.~J.~D.} \textbf{\bibinfo{volume}{32}},
  \bibinfo{pages}{149} (\bibinfo{year}{2004}).

\bibitem[{\citenamefont{{Stwalley} and {Wang}}(1999)}]{stwalley99a}
\bibinfo{author}{\bibfnamefont{W.~C.} \bibnamefont{{Stwalley}}}
  \bibnamefont{and} \bibinfo{author}{\bibfnamefont{H.}~\bibnamefont{{Wang}}},
  \bibinfo{journal}{J.~Mol.~Spectroscopy} \textbf{\bibinfo{volume}{195}},
  \bibinfo{pages}{194} (\bibinfo{year}{1999}).

\bibitem[{\citenamefont{Jones et~al.}(2006)\citenamefont{Jones, Tiesinga, Lett,
  and Julienne}}]{jones06a}
\bibinfo{author}{\bibfnamefont{K.~M.} \bibnamefont{Jones}},
  \bibinfo{author}{\bibfnamefont{E.}~\bibnamefont{Tiesinga}},
  \bibinfo{author}{\bibfnamefont{P.~D.} \bibnamefont{Lett}}, \bibnamefont{and}
  \bibinfo{author}{\bibfnamefont{P.~S.} \bibnamefont{Julienne}},
  \bibinfo{journal}{Rev.~Mod.~Phys.} \textbf{\bibinfo{volume}{78}},
  \bibinfo{pages}{483} (\bibinfo{year}{2006}).

\bibitem[{\citenamefont{Kohler et~al.}(2006)\citenamefont{Kohler, Goral, and
  Julienne}}]{kohler06a}
\bibinfo{author}{\bibfnamefont{T.}~\bibnamefont{Kohler}},
  \bibinfo{author}{\bibfnamefont{K.}~\bibnamefont{Goral}}, \bibnamefont{and}
  \bibinfo{author}{\bibfnamefont{P.~S.} \bibnamefont{Julienne}},
  \bibinfo{journal}{Rev.~Mod.~Phys.} \textbf{\bibinfo{volume}{78}},
  \bibinfo{pages}{1311} (\bibinfo{year}{2006}).

\bibitem[{\citenamefont{{Brue} and {Hutson}}(2012)}]{brue12a}
\bibinfo{author}{\bibfnamefont{D.~A.} \bibnamefont{{Brue}}} \bibnamefont{and}
  \bibinfo{author}{\bibfnamefont{J.~M.} \bibnamefont{{Hutson}}},
  \bibinfo{journal}{Phys.~Rev.~Lett.} \textbf{\bibinfo{volume}{108}},
  \bibinfo{eid}{043201} (\bibinfo{year}{2012}).

\bibitem[{\citenamefont{{Fioretti} et~al.}(1998)\citenamefont{{Fioretti},
  {Comparat}, {Crubellier}, {Dulieu}, {Masnou-Seeuws}, and
  {Pillet}}}]{fioretti98a}
\bibinfo{author}{\bibfnamefont{A.}~\bibnamefont{{Fioretti}}},
  \bibinfo{author}{\bibfnamefont{D.}~\bibnamefont{{Comparat}}},
  \bibinfo{author}{\bibfnamefont{A.}~\bibnamefont{{Crubellier}}},
  \bibinfo{author}{\bibfnamefont{O.}~\bibnamefont{{Dulieu}}},
  \bibinfo{author}{\bibfnamefont{F.}~\bibnamefont{{Masnou-Seeuws}}},
  \bibnamefont{and} \bibinfo{author}{\bibfnamefont{P.}~\bibnamefont{{Pillet}}},
  \bibinfo{journal}{Phys.Rev.~Lett.} \textbf{\bibinfo{volume}{80}},
  \bibinfo{pages}{4402} (\bibinfo{year}{1998}).

\bibitem[{\citenamefont{{Kraft} et~al.}(2006)\citenamefont{{Kraft}, {Staanum},
  {Lange}, {Vogel}, {Wester}, and {Weidem{\"u}ller}}}]{kraft06a}
\bibinfo{author}{\bibfnamefont{S.~D.} \bibnamefont{{Kraft}}},
  \bibinfo{author}{\bibfnamefont{P.}~\bibnamefont{{Staanum}}},
  \bibinfo{author}{\bibfnamefont{J.}~\bibnamefont{{Lange}}},
  \bibinfo{author}{\bibfnamefont{L.}~\bibnamefont{{Vogel}}},
  \bibinfo{author}{\bibfnamefont{R.}~\bibnamefont{{Wester}}}, \bibnamefont{and}
  \bibinfo{author}{\bibfnamefont{M.}~\bibnamefont{{Weidem{\"u}ller}}},
  \bibinfo{journal}{J.~Phys.~B} \textbf{\bibinfo{volume}{39}},
  \bibinfo{pages}{993} (\bibinfo{year}{2006}).

\bibitem[{\citenamefont{{Lozeille} et~al.}(2006)\citenamefont{{Lozeille},
  {Fioretti}, {Gabbanini}, {Huang}, {Pechkis}, {Wang}, {Gould}, {Eyler},
  {Stwalley}, {Aymar} et~al.}}]{lozeille06a}
\bibinfo{author}{\bibfnamefont{J.}~\bibnamefont{{Lozeille}}},
  \bibinfo{author}{\bibfnamefont{A.}~\bibnamefont{{Fioretti}}},
  \bibinfo{author}{\bibfnamefont{C.}~\bibnamefont{{Gabbanini}}},
  \bibinfo{author}{\bibfnamefont{Y.}~\bibnamefont{{Huang}}},
  \bibinfo{author}{\bibfnamefont{H.~K.} \bibnamefont{{Pechkis}}},
  \bibinfo{author}{\bibfnamefont{D.}~\bibnamefont{{Wang}}},
  \bibinfo{author}{\bibfnamefont{P.~L.} \bibnamefont{{Gould}}},
  \bibinfo{author}{\bibfnamefont{E.~E.} \bibnamefont{{Eyler}}},
  \bibinfo{author}{\bibfnamefont{W.~C.} \bibnamefont{{Stwalley}}},
  \bibinfo{author}{\bibfnamefont{M.}~\bibnamefont{{Aymar}}},
  \bibnamefont{et~al.}, \bibinfo{journal}{Eur.~Phys.~J.~D}
  \textbf{\bibinfo{volume}{39}}, \bibinfo{pages}{261} (\bibinfo{year}{2006}).

\bibitem[{\citenamefont{Zwierlein et~al.}(2003)\citenamefont{Zwierlein, Stan,
  Schunck, Raupach, Gupta, Hadzibabic, and Ketterle}}]{zwierlein03a}
\bibinfo{author}{\bibfnamefont{M.~W.} \bibnamefont{Zwierlein}},
  \bibinfo{author}{\bibfnamefont{C.~A.} \bibnamefont{Stan}},
  \bibinfo{author}{\bibfnamefont{C.~H.} \bibnamefont{Schunck}},
  \bibinfo{author}{\bibfnamefont{S.~M.~F.} \bibnamefont{Raupach}},
  \bibinfo{author}{\bibfnamefont{S.}~\bibnamefont{Gupta}},
  \bibinfo{author}{\bibfnamefont{Z.}~\bibnamefont{Hadzibabic}},
  \bibnamefont{and} \bibinfo{author}{\bibfnamefont{W.}~\bibnamefont{Ketterle}},
  \bibinfo{journal}{Phys. Rev. Lett.} \textbf{\bibinfo{volume}{91}},
  \bibinfo{pages}{250401} (\bibinfo{year}{2003}).

\bibitem[{\citenamefont{{Tscherbul} and {Krems}}(2006)}]{tscherbul06a}
\bibinfo{author}{\bibfnamefont{T.~V.} \bibnamefont{{Tscherbul}}}
  \bibnamefont{and} \bibinfo{author}{\bibfnamefont{R.~V.}
  \bibnamefont{{Krems}}}, \bibinfo{journal}{Phys.~Rev.~Lett.}
  \textbf{\bibinfo{volume}{97}}, \bibinfo{eid}{083201} (\bibinfo{year}{2006}),
  \eprint{arXiv:physics/0606027}.

\bibitem[{\citenamefont{{Kajita} et~al.}(2013)\citenamefont{{Kajita},
  {Gopakumar}, {Abe}, and {Hada}}}]{kajita13a}
\bibinfo{author}{\bibfnamefont{M.}~\bibnamefont{{Kajita}}},
  \bibinfo{author}{\bibfnamefont{G.}~\bibnamefont{{Gopakumar}}},
  \bibinfo{author}{\bibfnamefont{M.}~\bibnamefont{{Abe}}}, \bibnamefont{and}
  \bibinfo{author}{\bibfnamefont{M.}~\bibnamefont{{Hada}}},
  \bibinfo{journal}{J.~Phys.~B} \textbf{\bibinfo{volume}{46}},
  \bibinfo{pages}{025001} (\bibinfo{year}{2013}).

\bibitem[{\citenamefont{{Schlachta} et~al.}(1990)\citenamefont{{Schlachta},
  {Fischer}, {Rosmus}, and {Bondybey}}}]{schlachta90a}
\bibinfo{author}{\bibfnamefont{R.}~\bibnamefont{{Schlachta}}},
  \bibinfo{author}{\bibfnamefont{I.}~\bibnamefont{{Fischer}}},
  \bibinfo{author}{\bibfnamefont{P.}~\bibnamefont{{Rosmus}}}, \bibnamefont{and}
  \bibinfo{author}{\bibfnamefont{V.~E.} \bibnamefont{{Bondybey}}},
  \bibinfo{journal}{Chem.~Phys.~Lett.} \textbf{\bibinfo{volume}{170}},
  \bibinfo{pages}{485} (\bibinfo{year}{1990}).

\bibitem[{\citenamefont{{Bauschlicher}
  et~al.}(1992)\citenamefont{{Bauschlicher}, {Langhoff}, and
  {Partridge}}}]{bauschlicher92a}
\bibinfo{author}{\bibfnamefont{C.~W.} \bibnamefont{{Bauschlicher}},
  \bibfnamefont{Jr.}}, \bibinfo{author}{\bibfnamefont{S.~R.}
  \bibnamefont{{Langhoff}}}, \bibnamefont{and}
  \bibinfo{author}{\bibfnamefont{H.}~\bibnamefont{{Partridge}}},
  \bibinfo{journal}{J.~Chem.~Phys.} \textbf{\bibinfo{volume}{96}},
  \bibinfo{pages}{1240} (\bibinfo{year}{1992}).

\bibitem[{\citenamefont{{Allouche} and
  {Aubert-Fr{\`e}con}}(1994)}]{allouche94a}
\bibinfo{author}{\bibfnamefont{A.~R.} \bibnamefont{{Allouche}}}
  \bibnamefont{and}
  \bibinfo{author}{\bibfnamefont{M.}~\bibnamefont{{Aubert-Fr{\`e}con}}},
  \bibinfo{journal}{Chem.~Phys.~Lett.} \textbf{\bibinfo{volume}{222}},
  \bibinfo{pages}{524} (\bibinfo{year}{1994}).

\bibitem[{\citenamefont{{Berry}}(1997)}]{berry97a}
\bibinfo{author}{\bibfnamefont{K.}~\bibnamefont{{Berry}}},
  \bibinfo{journal}{Chem.~Phys.~Lett.} \textbf{\bibinfo{volume}{279}},
  \bibinfo{pages}{44} (\bibinfo{year}{1997}).

\bibitem[{\citenamefont{{Bruna} and {Grein}}(2002)}]{bruna02a}
\bibinfo{author}{\bibfnamefont{P.~J.} \bibnamefont{{Bruna}}} \bibnamefont{and}
  \bibinfo{author}{\bibfnamefont{F.}~\bibnamefont{{Grein}}},
  \bibinfo{journal}{Mol.~Phys.~} \textbf{\bibinfo{volume}{100}},
  \bibinfo{pages}{1681} (\bibinfo{year}{2002}).

\bibitem[{\citenamefont{{Ivanova} et~al.}(2011)\citenamefont{{Ivanova},
  {Stein}, {Pashov}, {Stolyarov}, {Kn{\"o}ckel}, and {Tiemann}}}]{ivanova11a}
\bibinfo{author}{\bibfnamefont{M.}~\bibnamefont{{Ivanova}}},
  \bibinfo{author}{\bibfnamefont{A.}~\bibnamefont{{Stein}}},
  \bibinfo{author}{\bibfnamefont{A.}~\bibnamefont{{Pashov}}},
  \bibinfo{author}{\bibfnamefont{A.~V.} \bibnamefont{{Stolyarov}}},
  \bibinfo{author}{\bibfnamefont{H.}~\bibnamefont{{Kn{\"o}ckel}}},
  \bibnamefont{and}
  \bibinfo{author}{\bibfnamefont{E.}~\bibnamefont{{Tiemann}}},
  \bibinfo{journal}{J.~Chem.~Phys.} \textbf{\bibinfo{volume}{135}},
  \bibinfo{pages}{174303} (\bibinfo{year}{2011}).

\bibitem[{\citenamefont{Guo et~al.}(2013)\citenamefont{Guo, Bajdich, Mitas, and
  Reynolds}}]{guo13a}
\bibinfo{author}{\bibfnamefont{S.}~\bibnamefont{Guo}},
  \bibinfo{author}{\bibfnamefont{M.}~\bibnamefont{Bajdich}},
  \bibinfo{author}{\bibfnamefont{L.}~\bibnamefont{Mitas}}, \bibnamefont{and}
  \bibinfo{author}{\bibfnamefont{P.~J.} \bibnamefont{Reynolds}},
  \bibinfo{journal}{ArXiv e-prints}  (\bibinfo{year}{2013}),
  \eprint{1301.1723}.

\bibitem[{\citenamefont{{Gopakumar} et~al.}(2010)\citenamefont{{Gopakumar},
  {Abe}, {Das}, {Hada}, and {Hirao}}}]{gopakumar10a}
\bibinfo{author}{\bibfnamefont{G.}~\bibnamefont{{Gopakumar}}},
  \bibinfo{author}{\bibfnamefont{M.}~\bibnamefont{{Abe}}},
  \bibinfo{author}{\bibfnamefont{B.~P.} \bibnamefont{{Das}}},
  \bibinfo{author}{\bibfnamefont{M.}~\bibnamefont{{Hada}}}, \bibnamefont{and}
  \bibinfo{author}{\bibfnamefont{K.}~\bibnamefont{{Hirao}}},
  \bibinfo{journal}{J.~Chem.~Phys.} \textbf{\bibinfo{volume}{133}},
  \bibinfo{pages}{124317} (\bibinfo{year}{2010}).

\bibitem[{\citenamefont{{Zhang} et~al.}(2010)\citenamefont{{Zhang},
  {Sadeghpour}, and {Dalgarno}}}]{zhang10a}
\bibinfo{author}{\bibfnamefont{P.}~\bibnamefont{{Zhang}}},
  \bibinfo{author}{\bibfnamefont{H.~R.} \bibnamefont{{Sadeghpour}}},
  \bibnamefont{and}
  \bibinfo{author}{\bibfnamefont{A.}~\bibnamefont{{Dalgarno}}},
  \bibinfo{journal}{J.~Chem.~Phys.} \textbf{\bibinfo{volume}{133}},
  \bibinfo{pages}{044306} (\bibinfo{year}{2010}).

\bibitem[{\citenamefont{{{\.Z}uchowski}
  et~al.}(2010)\citenamefont{{{\.Z}uchowski}, {Aldegunde}, and
  {Hutson}}}]{zuchowski10a}
\bibinfo{author}{\bibfnamefont{P.~S.} \bibnamefont{{{\.Z}uchowski}}},
  \bibinfo{author}{\bibfnamefont{J.}~\bibnamefont{{Aldegunde}}},
  \bibnamefont{and} \bibinfo{author}{\bibfnamefont{J.~M.}
  \bibnamefont{{Hutson}}}, \bibinfo{journal}{Phys.~Rev.~Lett.}
  \textbf{\bibinfo{volume}{105}}, \bibinfo{eid}{153201} (\bibinfo{year}{2010}).

\bibitem[{\citenamefont{{S{\o}rensen} et~al.}(2009)\citenamefont{{S{\o}rensen},
  {Knecht}, {Fleig}, and {Marian}}}]{sorensen09a}
\bibinfo{author}{\bibfnamefont{L.~K.} \bibnamefont{{S{\o}rensen}}},
  \bibinfo{author}{\bibfnamefont{S.}~\bibnamefont{{Knecht}}},
  \bibinfo{author}{\bibfnamefont{T.}~\bibnamefont{{Fleig}}}, \bibnamefont{and}
  \bibinfo{author}{\bibfnamefont{C.~M.} \bibnamefont{{Marian}}},
  \bibinfo{journal}{J.~Phys.~Chem.~A} \textbf{\bibinfo{volume}{113}},
  \bibinfo{pages}{12607} (\bibinfo{year}{2009}).

\bibitem[{\citenamefont{{M{\"u}nchow} et~al.}(2011)\citenamefont{{M{\"u}nchow},
  {Bruni}, {Madalinski}, and {G{\"o}rlitz}}}]{munchow11a}
\bibinfo{author}{\bibfnamefont{F.}~\bibnamefont{{M{\"u}nchow}}},
  \bibinfo{author}{\bibfnamefont{C.}~\bibnamefont{{Bruni}}},
  \bibinfo{author}{\bibfnamefont{M.}~\bibnamefont{{Madalinski}}},
  \bibnamefont{and}
  \bibinfo{author}{\bibfnamefont{A.}~\bibnamefont{{G{\"o}rlitz}}},
  \bibinfo{journal}{Physical Chemistry Chemical Physics (Incorporating Faraday
  Transactions)} \textbf{\bibinfo{volume}{13}}, \bibinfo{pages}{18734}
  (\bibinfo{year}{2011}).

\bibitem[{\citenamefont{Lamb et~al.}(2012)\citenamefont{Lamb, McCann,
  McLaughlin, Goold, Wells, and Lane}}]{lamb12a}
\bibinfo{author}{\bibfnamefont{H.~D.~L.} \bibnamefont{Lamb}},
  \bibinfo{author}{\bibfnamefont{J.~F.} \bibnamefont{McCann}},
  \bibinfo{author}{\bibfnamefont{B.~M.} \bibnamefont{McLaughlin}},
  \bibinfo{author}{\bibfnamefont{J.}~\bibnamefont{Goold}},
  \bibinfo{author}{\bibfnamefont{N.}~\bibnamefont{Wells}}, \bibnamefont{and}
  \bibinfo{author}{\bibfnamefont{I.}~\bibnamefont{Lane}},
  \bibinfo{journal}{Phys. Rev. A} \textbf{\bibinfo{volume}{86}},
  \bibinfo{pages}{022716} (\bibinfo{year}{2012}).

\bibitem[{\citenamefont{{Kotochigova} et~al.}(2011)\citenamefont{{Kotochigova},
  {Petrov}, {Linnik}, {K{\l}os}, and {Julienne}}}]{kotochigova11a}
\bibinfo{author}{\bibfnamefont{S.}~\bibnamefont{{Kotochigova}}},
  \bibinfo{author}{\bibfnamefont{A.}~\bibnamefont{{Petrov}}},
  \bibinfo{author}{\bibfnamefont{M.}~\bibnamefont{{Linnik}}},
  \bibinfo{author}{\bibfnamefont{J.}~\bibnamefont{{K{\l}os}}},
  \bibnamefont{and} \bibinfo{author}{\bibfnamefont{P.~S.}
  \bibnamefont{{Julienne}}}, \bibinfo{journal}{J.~Chem.~Phys.}
  \textbf{\bibinfo{volume}{135}}, \bibinfo{pages}{164108}
  (\bibinfo{year}{2011}), \eprint{1108.3530}.

\bibitem[{\citenamefont{{Gopakumar} et~al.}(2011)\citenamefont{{Gopakumar},
  {Abe}, {Kajita}, and {Hada}}}]{gopakumar11a}
\bibinfo{author}{\bibfnamefont{G.}~\bibnamefont{{Gopakumar}}},
  \bibinfo{author}{\bibfnamefont{M.}~\bibnamefont{{Abe}}},
  \bibinfo{author}{\bibfnamefont{M.}~\bibnamefont{{Kajita}}}, \bibnamefont{and}
  \bibinfo{author}{\bibfnamefont{M.}~\bibnamefont{{Hada}}},
  \bibinfo{journal}{Phys.~Rev.~A} \textbf{\bibinfo{volume}{84}},
  \bibinfo{eid}{062514} (\bibinfo{year}{2011}).

\bibitem[{\citenamefont{Dalgarno and Davison}(1966)}]{dalgarno66a}
\bibinfo{author}{\bibfnamefont{A.}~\bibnamefont{Dalgarno}} \bibnamefont{and}
  \bibinfo{author}{\bibfnamefont{W.~D.} \bibnamefont{Davison}},
  \bibinfo{journal}{Adv.~At.~Mol.~Phys.} \textbf{\bibinfo{volume}{2}},
  \bibinfo{pages}{1} (\bibinfo{year}{1966}).

\bibitem[{\citenamefont{Dalgarno}(1967)}]{dalgarno67a}
\bibinfo{author}{\bibfnamefont{A.}~\bibnamefont{Dalgarno}},
  \bibinfo{journal}{Adv.~Chem.~Phys.} \textbf{\bibinfo{volume}{12}},
  \bibinfo{pages}{143} (\bibinfo{year}{1967}).

\bibitem[{\citenamefont{Zhang and Mitroy}(2007)}]{zhang07c}
\bibinfo{author}{\bibfnamefont{J.~Y.} \bibnamefont{Zhang}} \bibnamefont{and}
  \bibinfo{author}{\bibfnamefont{J.}~\bibnamefont{Mitroy}},
  \bibinfo{journal}{Phys.~Rev.~A} \textbf{\bibinfo{volume}{76}},
  \bibinfo{pages}{022705} (\bibinfo{year}{2007}).

\bibitem[{\citenamefont{Bukta and Meath}(1974)}]{bukta74a}
\bibinfo{author}{\bibfnamefont{J.~F.} \bibnamefont{Bukta}} \bibnamefont{and}
  \bibinfo{author}{\bibfnamefont{W.~J.} \bibnamefont{Meath}},
  \bibinfo{journal}{Mol.~Phys.} \textbf{\bibinfo{volume}{27}},
  \bibinfo{pages}{1235} (\bibinfo{year}{1974}).

\bibitem[{\citenamefont{Mitroy and Bromley}(2003{\natexlab{a}})}]{mitroy03f}
\bibinfo{author}{\bibfnamefont{J.}~\bibnamefont{Mitroy}} \bibnamefont{and}
  \bibinfo{author}{\bibfnamefont{M.~W.~J.} \bibnamefont{Bromley}},
  \bibinfo{journal}{Phys.~Rev.~A} \textbf{\bibinfo{volume}{68}},
  \bibinfo{pages}{052714} (\bibinfo{year}{2003}{\natexlab{a}}).

\bibitem[{\citenamefont{Mitroy and Zhang}(2007)}]{mitroy07e}
\bibinfo{author}{\bibfnamefont{J.}~\bibnamefont{Mitroy}} \bibnamefont{and}
  \bibinfo{author}{\bibfnamefont{J.~Y.} \bibnamefont{Zhang}},
  \bibinfo{journal}{Phys.~Rev.~A} \textbf{\bibinfo{volume}{76}},
  \bibinfo{pages}{062703} (\bibinfo{year}{2007}).

\bibitem[{\citenamefont{Hameed et~al.}(1968)\citenamefont{Hameed, Herzenberg,
  and James}}]{hameed68a}
\bibinfo{author}{\bibfnamefont{S.}~\bibnamefont{Hameed}},
  \bibinfo{author}{\bibfnamefont{A.}~\bibnamefont{Herzenberg}},
  \bibnamefont{and} \bibinfo{author}{\bibfnamefont{M.~G.} \bibnamefont{James}},
  \bibinfo{journal}{J.~Phys.~B} \textbf{\bibinfo{volume}{1}},
  \bibinfo{pages}{822} (\bibinfo{year}{1968}).

\bibitem[{\citenamefont{Hameed}(1972)}]{hameed72a}
\bibinfo{author}{\bibfnamefont{S.}~\bibnamefont{Hameed}},
  \bibinfo{journal}{J.~Phys.~B} \textbf{\bibinfo{volume}{5}},
  \bibinfo{pages}{746} (\bibinfo{year}{1972}).

\bibitem[{\citenamefont{Vaeck et~al.}(1992)\citenamefont{Vaeck, Godefroid, and
  Froese~Fischer}}]{vaeck92a}
\bibinfo{author}{\bibfnamefont{N.}~\bibnamefont{Vaeck}},
  \bibinfo{author}{\bibfnamefont{M.}~\bibnamefont{Godefroid}},
  \bibnamefont{and}
  \bibinfo{author}{\bibfnamefont{C.}~\bibnamefont{Froese~Fischer}},
  \bibinfo{journal}{Phys. Rev. A} \textbf{\bibinfo{volume}{46}},
  \bibinfo{pages}{3704} (\bibinfo{year}{1992}).

\bibitem[{\citenamefont{Mitroy}(1993)}]{mitroy93a}
\bibinfo{author}{\bibfnamefont{J.}~\bibnamefont{Mitroy}},
  \bibinfo{journal}{J.~Phys.~B} \textbf{\bibinfo{volume}{26}},
  \bibinfo{pages}{2201} (\bibinfo{year}{1993}).

\bibitem[{\citenamefont{Mitroy and Bromley}(2003{\natexlab{b}})}]{mitroy03e}
\bibinfo{author}{\bibfnamefont{J.}~\bibnamefont{Mitroy}} \bibnamefont{and}
  \bibinfo{author}{\bibfnamefont{M.~W.~J.} \bibnamefont{Bromley}},
  \bibinfo{journal}{Phys.~Rev.~A} \textbf{\bibinfo{volume}{68}},
  \bibinfo{pages}{035201} (\bibinfo{year}{2003}{\natexlab{b}}).

\bibitem[{\citenamefont{Mitroy and Bromley}(2004)}]{mitroy04b}
\bibinfo{author}{\bibfnamefont{J.}~\bibnamefont{Mitroy}} \bibnamefont{and}
  \bibinfo{author}{\bibfnamefont{M.~W.~J.} \bibnamefont{Bromley}},
  \bibinfo{journal}{Phys.~Rev.~A} \textbf{\bibinfo{volume}{70}},
  \bibinfo{pages}{052503} (\bibinfo{year}{2004}).

\bibitem[{\citenamefont{Zhang et~al.}(2007)\citenamefont{Zhang, Mitroy, and
  Bromley}}]{zhang07a}
\bibinfo{author}{\bibfnamefont{J.~Y.} \bibnamefont{Zhang}},
  \bibinfo{author}{\bibfnamefont{J.}~\bibnamefont{Mitroy}}, \bibnamefont{and}
  \bibinfo{author}{\bibfnamefont{M.~W.~J.} \bibnamefont{Bromley}},
  \bibinfo{journal}{Phys.~Rev.~A} \textbf{\bibinfo{volume}{75}},
  \bibinfo{pages}{042509} (\bibinfo{year}{2007}).

\bibitem[{\citenamefont{Kumar and Meath}(1985)}]{kumar85a}
\bibinfo{author}{\bibfnamefont{A.}~\bibnamefont{Kumar}} \bibnamefont{and}
  \bibinfo{author}{\bibfnamefont{W.~J.} \bibnamefont{Meath}},
  \bibinfo{journal}{Mol. Phys.} \textbf{\bibinfo{volume}{54}},
  \bibinfo{pages}{823} (\bibinfo{year}{1985}).

\bibitem[{\citenamefont{Mitroy et~al.}(2010)\citenamefont{Mitroy, Safronova,
  and Clark}}]{mitroy10a}
\bibinfo{author}{\bibfnamefont{J.}~\bibnamefont{Mitroy}},
  \bibinfo{author}{\bibfnamefont{M.~S.} \bibnamefont{Safronova}},
  \bibnamefont{and} \bibinfo{author}{\bibfnamefont{C.~W.} \bibnamefont{Clark}},
  \bibinfo{journal}{J.~Phys.~B} \textbf{\bibinfo{volume}{43}},
  \bibinfo{pages}{202001} (\bibinfo{year}{2010}).

\bibitem[{\citenamefont{{Johnson} et~al.}(2008)\citenamefont{{Johnson},
  {Safronova}, {Derevianko}, and {Safronova}}}]{johnson08a}
\bibinfo{author}{\bibfnamefont{W.~R.} \bibnamefont{{Johnson}}},
  \bibinfo{author}{\bibfnamefont{U.~I.} \bibnamefont{{Safronova}}},
  \bibinfo{author}{\bibfnamefont{A.}~\bibnamefont{{Derevianko}}},
  \bibnamefont{and} \bibinfo{author}{\bibfnamefont{M.~S.}
  \bibnamefont{{Safronova}}}, \bibinfo{journal}{Phys.~Rev.~A}
  \textbf{\bibinfo{volume}{77}}, \bibinfo{pages}{022510}
  (\bibinfo{year}{2008}).

\bibitem[{\citenamefont{{Tang} et~al.}(2010)\citenamefont{{Tang}, {Yan}, {Shi},
  and {Mitroy}}}]{tang10a}
\bibinfo{author}{\bibfnamefont{L.-Y.} \bibnamefont{{Tang}}},
  \bibinfo{author}{\bibfnamefont{Z.-C.} \bibnamefont{{Yan}}},
  \bibinfo{author}{\bibfnamefont{T.-Y.} \bibnamefont{{Shi}}}, \bibnamefont{and}
  \bibinfo{author}{\bibfnamefont{J.}~\bibnamefont{{Mitroy}}},
  \bibinfo{journal}{Phys.~Rev.~A} \textbf{\bibinfo{volume}{81}},
  \bibinfo{pages}{042521} (\bibinfo{year}{2010}).

\bibitem[{\citenamefont{Miffre et~al.}(2006)\citenamefont{Miffre, Jacquet,
  Buchner, Trenec, and Vigue}}]{miffre06a}
\bibinfo{author}{\bibfnamefont{A.}~\bibnamefont{Miffre}},
  \bibinfo{author}{\bibfnamefont{M.}~\bibnamefont{Jacquet}},
  \bibinfo{author}{\bibfnamefont{M.}~\bibnamefont{Buchner}},
  \bibinfo{author}{\bibfnamefont{G.}~\bibnamefont{Trenec}}, \bibnamefont{and}
  \bibinfo{author}{\bibfnamefont{J.}~\bibnamefont{Vigue}},
  \bibinfo{journal}{Eur.~Phys.~J.~D} \textbf{\bibinfo{volume}{38}},
  \bibinfo{pages}{353} (\bibinfo{year}{2006}).

\bibitem[{\citenamefont{Derevianko et~al.}(1999)\citenamefont{Derevianko,
  Johnson, Safronova, and Babb}}]{derevianko99a}
\bibinfo{author}{\bibfnamefont{A.}~\bibnamefont{Derevianko}},
  \bibinfo{author}{\bibfnamefont{W.~R.} \bibnamefont{Johnson}},
  \bibinfo{author}{\bibfnamefont{M.~S.} \bibnamefont{Safronova}},
  \bibnamefont{and} \bibinfo{author}{\bibfnamefont{J.~F.} \bibnamefont{Babb}},
  \bibinfo{journal}{Phys.~Rev.~Lett.} \textbf{\bibinfo{volume}{82}},
  \bibinfo{pages}{3589} (\bibinfo{year}{1999}).

\bibitem[{\citenamefont{Ekstrom et~al.}(1995)\citenamefont{Ekstrom,
  Schmiedmayer, Chapman, Hammond, and Pritchard}}]{ekstrom95a}
\bibinfo{author}{\bibfnamefont{C.~R.} \bibnamefont{Ekstrom}},
  \bibinfo{author}{\bibfnamefont{J.}~\bibnamefont{Schmiedmayer}},
  \bibinfo{author}{\bibfnamefont{M.~S.} \bibnamefont{Chapman}},
  \bibinfo{author}{\bibfnamefont{T.~D.} \bibnamefont{Hammond}},
  \bibnamefont{and} \bibinfo{author}{\bibfnamefont{D.~E.}
  \bibnamefont{Pritchard}}, \bibinfo{journal}{Phys. Rev. A}
  \textbf{\bibinfo{volume}{51}}, \bibinfo{pages}{3883} (\bibinfo{year}{1995}).

\bibitem[{\citenamefont{Holmgren et~al.}(2010)\citenamefont{Holmgren, Revelle,
  Lonij, and Cronin}}]{holmgren10a}
\bibinfo{author}{\bibfnamefont{W.~F.} \bibnamefont{Holmgren}},
  \bibinfo{author}{\bibfnamefont{M.~C.} \bibnamefont{Revelle}},
  \bibinfo{author}{\bibfnamefont{V.~P.~A.} \bibnamefont{Lonij}},
  \bibnamefont{and} \bibinfo{author}{\bibfnamefont{A.~D.}
  \bibnamefont{Cronin}}, \bibinfo{journal}{Phys.~Rev.~A}
  \textbf{\bibinfo{volume}{81}}, \bibinfo{pages}{053607}
  (\bibinfo{year}{2010}).

\bibitem[{\citenamefont{Porsev and Derevianko}(2003)}]{porsev03a}
\bibinfo{author}{\bibfnamefont{S.~G.} \bibnamefont{Porsev}} \bibnamefont{and}
  \bibinfo{author}{\bibfnamefont{A.}~\bibnamefont{Derevianko}},
  \bibinfo{journal}{J.~Chem.~Phys.} \textbf{\bibinfo{volume}{119}},
  \bibinfo{pages}{844} (\bibinfo{year}{2003}).

\bibitem[{\citenamefont{{Tang} et~al.}(2009)\citenamefont{{Tang}, {Yan}, {Shi},
  and {Babb}}}]{tang09a}
\bibinfo{author}{\bibfnamefont{L.-Y.} \bibnamefont{{Tang}}},
  \bibinfo{author}{\bibfnamefont{Z.-C.} \bibnamefont{{Yan}}},
  \bibinfo{author}{\bibfnamefont{T.-Y.} \bibnamefont{{Shi}}}, \bibnamefont{and}
  \bibinfo{author}{\bibfnamefont{J.~F.} \bibnamefont{{Babb}}},
  \bibinfo{journal}{Phys.~Rev.~A} \textbf{\bibinfo{volume}{79}},
  \bibinfo{pages}{062712} (\bibinfo{year}{2009}).

\bibitem[{\citenamefont{{Safronova} and {Safronova}}(2008)}]{safronova08b}
\bibinfo{author}{\bibfnamefont{U.~I.} \bibnamefont{{Safronova}}}
  \bibnamefont{and} \bibinfo{author}{\bibfnamefont{M.~S.}
  \bibnamefont{{Safronova}}}, \bibinfo{journal}{Phys.~Rev.~A}
  \textbf{\bibinfo{volume}{78}}, \bibinfo{pages}{052504}
  (\bibinfo{year}{2008}).

\bibitem[{\citenamefont{Porsev and Derevianko}(2006)}]{porsev06a}
\bibinfo{author}{\bibfnamefont{S.~G.} \bibnamefont{Porsev}} \bibnamefont{and}
  \bibinfo{author}{\bibfnamefont{A.}~\bibnamefont{Derevianko}},
  \bibinfo{journal}{JETP} \textbf{\bibinfo{volume}{102}}, \bibinfo{pages}{195}
  (\bibinfo{year}{2006}).

\bibitem[{\citenamefont{Komasa}(2002)}]{komasa02a}
\bibinfo{author}{\bibfnamefont{J.}~\bibnamefont{Komasa}},
  \bibinfo{journal}{Phys.~Rev.~A} \textbf{\bibinfo{volume}{65}},
  \bibinfo{pages}{012506} (\bibinfo{year}{2002}).

\bibitem[{\citenamefont{Lundin et~al.}(1973)\citenamefont{Lundin, Engman,
  Hilke, and Martinson}}]{lundin73a}
\bibinfo{author}{\bibfnamefont{L.}~\bibnamefont{Lundin}},
  \bibinfo{author}{\bibfnamefont{B.}~\bibnamefont{Engman}},
  \bibinfo{author}{\bibfnamefont{J.}~\bibnamefont{Hilke}}, \bibnamefont{and}
  \bibinfo{author}{\bibfnamefont{I.}~\bibnamefont{Martinson}},
  \bibinfo{journal}{Phys.~Scr.} \textbf{\bibinfo{volume}{8}},
  \bibinfo{pages}{274} (\bibinfo{year}{1973}).

\bibitem[{\citenamefont{Miller}(1995)}]{miller95b}
\bibinfo{author}{\bibfnamefont{T.~M.} \bibnamefont{Miller}},
  \emph{\bibinfo{title}{Atomic and Molecular Polarizabilities}}
  (\bibinfo{publisher}{CRC Press}, \bibinfo{address}{Boca Raton, Florida},
  \bibinfo{year}{1995}), vol.~\bibinfo{volume}{76},
  chap.~\bibinfo{chapter}{10}, pp. \bibinfo{pages}{10--192}.

\bibitem[{\citenamefont{Safronova et~al.}(2013)\citenamefont{Safronova, Porsev,
  Safronova, Kozlov, and Clark}}]{safronova13b}
\bibinfo{author}{\bibfnamefont{M.~S.} \bibnamefont{Safronova}},
  \bibinfo{author}{\bibfnamefont{S.~G.} \bibnamefont{Porsev}},
  \bibinfo{author}{\bibfnamefont{U.~I.} \bibnamefont{Safronova}},
  \bibinfo{author}{\bibfnamefont{M.~G.} \bibnamefont{Kozlov}},
  \bibnamefont{and} \bibinfo{author}{\bibfnamefont{C.~W.} \bibnamefont{Clark}},
  \bibinfo{journal}{Phys. Rev. A} \textbf{\bibinfo{volume}{87}},
  \bibinfo{pages}{012509} (\bibinfo{year}{2013}).

\bibitem[{\citenamefont{Mitroy et~al.}(1988)\citenamefont{Mitroy, Griffin,
  Norcross, and Pindzola}}]{mitroy88d}
\bibinfo{author}{\bibfnamefont{J.}~\bibnamefont{Mitroy}},
  \bibinfo{author}{\bibfnamefont{D.~C.} \bibnamefont{Griffin}},
  \bibinfo{author}{\bibfnamefont{D.~W.} \bibnamefont{Norcross}},
  \bibnamefont{and} \bibinfo{author}{\bibfnamefont{M.~S.}
  \bibnamefont{Pindzola}}, \bibinfo{journal}{Phys.~Rev.~A}
  \textbf{\bibinfo{volume}{38}}, \bibinfo{pages}{3339} (\bibinfo{year}{1988}).

\bibitem[{\citenamefont{{Mitroy}}(2010)}]{mitroy10d}
\bibinfo{author}{\bibfnamefont{J.}~\bibnamefont{{Mitroy}}},
  \bibinfo{journal}{Phys.~Rev.~A} \textbf{\bibinfo{volume}{82}},
  \bibinfo{pages}{052516} (\bibinfo{year}{2010}).

\bibitem[{\citenamefont{Mitroy and Zhang}(2008{\natexlab{a}})}]{mitroy08a}
\bibinfo{author}{\bibfnamefont{J.}~\bibnamefont{Mitroy}} \bibnamefont{and}
  \bibinfo{author}{\bibfnamefont{J.~Y.} \bibnamefont{Zhang}},
  \bibinfo{journal}{Mol.~Phys.} \textbf{\bibinfo{volume}{106}},
  \bibinfo{pages}{127} (\bibinfo{year}{2008}{\natexlab{a}}).

\bibitem[{\citenamefont{Mitroy and Zhang}(2008{\natexlab{b}})}]{mitroy08g}
\bibinfo{author}{\bibfnamefont{J.}~\bibnamefont{Mitroy}} \bibnamefont{and}
  \bibinfo{author}{\bibfnamefont{J.~Y.} \bibnamefont{Zhang}},
  \bibinfo{journal}{J.~Chem.~Phys.} \textbf{\bibinfo{volume}{128}},
  \bibinfo{pages}{134305} (\bibinfo{year}{2008}{\natexlab{b}}).

\bibitem[{\citenamefont{Mitroy and Zhang}(2010)}]{mitroy10b}
\bibinfo{author}{\bibfnamefont{J.}~\bibnamefont{Mitroy}} \bibnamefont{and}
  \bibinfo{author}{\bibfnamefont{J.~Y.} \bibnamefont{Zhang}},
  \bibinfo{journal}{Mol.~Phys.} \textbf{\bibinfo{volume}{108}},
  \bibinfo{pages}{1999} (\bibinfo{year}{2010}).

\bibitem[{\citenamefont{{Yasuda} et~al.}(2006)\citenamefont{{Yasuda},
  {Kishimoto}, {Takamoto}, and {Katori}}}]{yasuda06a}
\bibinfo{author}{\bibfnamefont{M.}~\bibnamefont{{Yasuda}}},
  \bibinfo{author}{\bibfnamefont{T.}~\bibnamefont{{Kishimoto}}},
  \bibinfo{author}{\bibfnamefont{M.}~\bibnamefont{{Takamoto}}},
  \bibnamefont{and} \bibinfo{author}{\bibfnamefont{H.}~\bibnamefont{{Katori}}},
  \bibinfo{journal}{Phys.~Rev.~A} \textbf{\bibinfo{volume}{73}},
  \bibinfo{pages}{011403} (\bibinfo{year}{2006}).

\bibitem[{\citenamefont{Miller and Bederson}(1977)}]{miller77a}
\bibinfo{author}{\bibfnamefont{T.~M.} \bibnamefont{Miller}} \bibnamefont{and}
  \bibinfo{author}{\bibfnamefont{B.}~\bibnamefont{Bederson}},
  \bibinfo{journal}{Adv.~At.~Mol.~Phys.} \textbf{\bibinfo{volume}{13}},
  \bibinfo{pages}{1} (\bibinfo{year}{1977}).

\bibitem[{\citenamefont{{Derevianko} et~al.}(2010)\citenamefont{{Derevianko},
  {Porsev}, and {Babb}}}]{derevianko10a}
\bibinfo{author}{\bibfnamefont{A.}~\bibnamefont{{Derevianko}}},
  \bibinfo{author}{\bibfnamefont{S.~G.} \bibnamefont{{Porsev}}},
  \bibnamefont{and} \bibinfo{author}{\bibfnamefont{J.~F.}
  \bibnamefont{{Babb}}}, \bibinfo{journal}{At.~Data Nucl.~Data Tables}
  \textbf{\bibinfo{volume}{96}}, \bibinfo{pages}{323} (\bibinfo{year}{2010}).

\bibitem[{\citenamefont{Safronova et~al.}(1999)\citenamefont{Safronova,
  Johnson, and Derevianko}}]{safronova99a}
\bibinfo{author}{\bibfnamefont{M.~S.} \bibnamefont{Safronova}},
  \bibinfo{author}{\bibfnamefont{W.~R.} \bibnamefont{Johnson}},
  \bibnamefont{and}
  \bibinfo{author}{\bibfnamefont{A.}~\bibnamefont{Derevianko}},
  \bibinfo{journal}{Phys.~Rev.~A} \textbf{\bibinfo{volume}{60}},
  \bibinfo{pages}{4476} (\bibinfo{year}{1999}).

\bibitem[{\citenamefont{Katori et~al.}(2003)\citenamefont{Katori, Takamoto,
  Pal'chikov, and Ovsiannikov}}]{katori03a}
\bibinfo{author}{\bibfnamefont{H.}~\bibnamefont{Katori}},
  \bibinfo{author}{\bibfnamefont{M.}~\bibnamefont{Takamoto}},
  \bibinfo{author}{\bibfnamefont{V.~G.} \bibnamefont{Pal'chikov}},
  \bibnamefont{and} \bibinfo{author}{\bibfnamefont{V.~D.}
  \bibnamefont{Ovsiannikov}}, \bibinfo{journal}{Phys.~Rev.~Lett.}
  \textbf{\bibinfo{volume}{91}}, \bibinfo{pages}{173005}
  (\bibinfo{year}{2003}).

\bibitem[{\citenamefont{Arora et~al.}(2007)\citenamefont{Arora, Safronova, and
  Clark}}]{arora07c}
\bibinfo{author}{\bibfnamefont{B.}~\bibnamefont{Arora}},
  \bibinfo{author}{\bibfnamefont{M.~S.} \bibnamefont{Safronova}},
  \bibnamefont{and} \bibinfo{author}{\bibfnamefont{C.~W.} \bibnamefont{Clark}},
  \bibinfo{journal}{Phys. Rev. A} \textbf{\bibinfo{volume}{76}},
  \bibinfo{pages}{052509} (\bibinfo{year}{2007}).

\bibitem[{\citenamefont{{Herold} et~al.}(2012)\citenamefont{{Herold}, {Vaidya},
  {Li}, {Rolston}, {Porto}, and {Safronova}}}]{herold12a}
\bibinfo{author}{\bibfnamefont{C.~D.} \bibnamefont{{Herold}}},
  \bibinfo{author}{\bibfnamefont{V.~D.} \bibnamefont{{Vaidya}}},
  \bibinfo{author}{\bibfnamefont{X.}~\bibnamefont{{Li}}},
  \bibinfo{author}{\bibfnamefont{S.~L.} \bibnamefont{{Rolston}}},
  \bibinfo{author}{\bibfnamefont{J.~V.} \bibnamefont{{Porto}}},
  \bibnamefont{and} \bibinfo{author}{\bibfnamefont{M.~S.}
  \bibnamefont{{Safronova}}}, \bibinfo{journal}{Phys.~Rev.~Lett.}
  \textbf{\bibinfo{volume}{109}}, \bibinfo{eid}{243003} (\bibinfo{year}{2012}).

\bibitem[{\citenamefont{{Holmgren} et~al.}(2012)\citenamefont{{Holmgren},
  {Trubko}, {Hromada}, and {Cronin}}}]{holmgren12a}
\bibinfo{author}{\bibfnamefont{W.~F.} \bibnamefont{{Holmgren}}},
  \bibinfo{author}{\bibfnamefont{R.}~\bibnamefont{{Trubko}}},
  \bibinfo{author}{\bibfnamefont{I.}~\bibnamefont{{Hromada}}},
  \bibnamefont{and} \bibinfo{author}{\bibfnamefont{A.~D.}
  \bibnamefont{{Cronin}}}, \bibinfo{journal}{Phys.~Rev.~Lett.}
  \textbf{\bibinfo{volume}{109}}, \bibinfo{eid}{243004} (\bibinfo{year}{2012}).

\bibitem[{\citenamefont{{Safronova} et~al.}(2012)\citenamefont{{Safronova},
  {Safronova}, and {Clark}}}]{safronova12c}
\bibinfo{author}{\bibfnamefont{M.~S.} \bibnamefont{{Safronova}}},
  \bibinfo{author}{\bibfnamefont{U.~I.} \bibnamefont{{Safronova}}},
  \bibnamefont{and} \bibinfo{author}{\bibfnamefont{C.~W.}
  \bibnamefont{{Clark}}}, \bibinfo{journal}{\pra}
  \textbf{\bibinfo{volume}{86}}, \bibinfo{eid}{042505} (\bibinfo{year}{2012}).

\bibitem[{\citenamefont{Jiang et~al.}(2013)\citenamefont{Jiang, Tang, and
  Mitroy}}]{jiang13a}
\bibinfo{author}{\bibfnamefont{J.}~\bibnamefont{Jiang}},
  \bibinfo{author}{\bibfnamefont{L.~Y.} \bibnamefont{Tang}}, \bibnamefont{and}
  \bibinfo{author}{\bibfnamefont{J.}~\bibnamefont{Mitroy}},
  \bibinfo{journal}{Phys. Rev. A} \textbf{\bibinfo{volume}{87}},
  \bibinfo{pages}{032518} (\bibinfo{year}{2013}).

\bibitem[{\citenamefont{Tang et~al.}(2013)\citenamefont{Tang, Bromley, Yan, and
  Mitroy}}]{tang13a}
\bibinfo{author}{\bibfnamefont{L.~Y.} \bibnamefont{Tang}},
  \bibinfo{author}{\bibfnamefont{M.~W.~J.} \bibnamefont{Bromley}},
  \bibinfo{author}{\bibfnamefont{Z.~C.} \bibnamefont{Yan}}, \bibnamefont{and}
  \bibinfo{author}{\bibfnamefont{J.}~\bibnamefont{Mitroy}},
  \bibinfo{journal}{Phys. Rev. A} \textbf{\bibinfo{volume}{87}},
  \bibinfo{pages}{032507} (\bibinfo{year}{2013}).

\bibitem[{\citenamefont{{Safronova} et~al.}(2013)\citenamefont{{Safronova},
  {Safronova}, and {Clark}}}]{safronova13a}
\bibinfo{author}{\bibfnamefont{M.~S.} \bibnamefont{{Safronova}}},
  \bibinfo{author}{\bibfnamefont{U.~I.} \bibnamefont{{Safronova}}},
  \bibnamefont{and} \bibinfo{author}{\bibfnamefont{C.~W.}
  \bibnamefont{{Clark}}}, \bibinfo{journal}{ArXiv e-prints}
  (\bibinfo{year}{2013}), \eprint{1301.3181}.

\bibitem[{\citenamefont{{Safronova} and {Safronova}}(2011)}]{safronova11b}
\bibinfo{author}{\bibfnamefont{M.~S.} \bibnamefont{{Safronova}}}
  \bibnamefont{and} \bibinfo{author}{\bibfnamefont{U.~I.}
  \bibnamefont{{Safronova}}}, \bibinfo{journal}{Phys.~Rev.~A}
  \textbf{\bibinfo{volume}{83}}, \bibinfo{eid}{052508} (\bibinfo{year}{2011}).

\bibitem[{\citenamefont{Standard and Certain}(1985)}]{standard85a}
\bibinfo{author}{\bibfnamefont{J.~M.} \bibnamefont{Standard}} \bibnamefont{and}
  \bibinfo{author}{\bibfnamefont{P.~R.} \bibnamefont{Certain}},
  \bibinfo{journal}{J.~Chem.~Phys.} \textbf{\bibinfo{volume}{83}},
  \bibinfo{pages}{3002} (\bibinfo{year}{1985}).

\bibitem[{\citenamefont{Marinescu and Sadeghpour}(1999)}]{marinescu99a}
\bibinfo{author}{\bibfnamefont{M.}~\bibnamefont{Marinescu}} \bibnamefont{and}
  \bibinfo{author}{\bibfnamefont{H.~R.} \bibnamefont{Sadeghpour}},
  \bibinfo{journal}{Phys.~Rev.~A} \textbf{\bibinfo{volume}{59}},
  \bibinfo{pages}{390} (\bibinfo{year}{1999}).

\end{thebibliography}

\end{document}